\documentclass[onecolumn,showpacs,aps,prd]{revtex4}
\def\OL{\overline} 

\usepackage{graphicx}
\usepackage{dcolumn}
\usepackage{amsmath}
\usepackage{epsfig}
\RequirePackage{xspace}

\begin{document}

\def\op{{\cal O}}
\def\lsim{\mathrel{\lower4pt\hbox{$\sim$}}\hskip-12pt\raise1.6pt\hbox{$<$}\;}
\def\gsim{\mathrel{\lower4pt\hbox{$\sim$}}\hskip-10pt\raise1.6pt\hbox{$>$}\;}
\def\Dd{\psi}
\def\pp{\lambda}
\def\ket{\rangle}
\def\BAR{\bar}
\def\xba{\bar}
\def\fa{{\cal A}}
\def\fm{{\cal M}}
\def\fl{{\cal L}}
\def\ufs{\Upsilon(5S)}
\def\ufour{\Upsilon(4S)}
\def\xcp{X_{CP}}
\def\ynotcp{Y}
\vspace*{-.5in}
\def\bfb{{\bf B}}
\def\fd{r_D}
\def\fb{r_B}
\def\hatA{\hat A}
\def\hatfd{{\hat r}_D}
\def\D{{\bf D}}
\def\pcc{(+ charge conjugate)}

\def\KS{K_S}
\def\KL{K_L}
\def\dmd{\Delta m_d}
\def\dms{\Delta m_s}
\def\dgd{\Delta \Gamma_d}
\def\dgs{\Delta \Gamma_s}

\def\uglu {\hskip 0pt plus 1fil minus 1fil} 
\def\uglux{\hskip 0pt plus .75fil minus .75fil}

\def\slashed#1{\setbox200=\hbox{$ #1 $}
    \hbox{\box200 \hskip -\wd200 \hbox to \wd200 {\uglu $/$ \uglux}}}

\def\slpar{\slashed\partial}
\def\sla{\slashed a}
\def\slb{\slashed b}
\def\slc{\slashed c}
\def\sld{\slashed d}
\def\sle{\slashed e}
\def\slf{\slashed f}
\def\slg{\slashed g}
\def\slh{\slashed h}
\def\sli{\slashed i}
\def\slj{\slashed j}
\def\slk{\slashed k}
\def\sll{\slashed l}
\def\slm{\slashed m}
\def\sln{\slashed n}
\def\slo{\slashed o}
\def\slp{\slashed p}
\def\slq{\slashed q}
\def\slr{\slashed r}
\def\sls{\slashed s}
\def\slt{\slashed t}
\def\slu{\slashed u}
\def\slv{\slashed v}
\def\slw{\slashed w}
\def\slx{\slashed x}
\def\sly{\slashed y}
\def\slz{\slashed z}
\def\slE{\slashed E}

\renewcommand{\Re}{\ensuremath{{\rm Re}}}
\renewcommand{\Im}{\ensuremath{{\rm Im}}}

\def\cals{{\cal S}}

%AS1
\title{
%%%%
%
\begin{flushright}
{KEK preprint 2004-56}\\
{BNL-HET-04/18~~~}\\
\end{flushright}
%
%%%%
\vskip 10mm
\large\bf
\boldmath
Mixing-Induced $CP$ Violation in $B \to P_1 P_2 \gamma$
in Search of Clean New Physics Signals
}

%AS1 end  
\author{David Atwood}
\affiliation{
  Dept. of Physics and Astronomy, Iowa State University, Ames, IA 50011
}
\author{Tim Gershon}
\affiliation{
  High Energy Accelerator Research Organization (KEK), Tsukuba, Ibaraki, Japan
}
\author{Masashi Hazumi}
\affiliation{
  High Energy Accelerator Research Organization (KEK), Tsukuba, Ibaraki, Japan
}
\author{Amarjit Soni}
\affiliation{
  Theory Group, Brookhaven National Laboratory, Upton, NY 11973
}

\date{\today}

\begin{abstract}

We show that in a decay of the form $B_d$ or $B_s \to P_1 P_2 \gamma$
(where $P_1$ and $P_2$ are pseudoscalar mesons), through a flavor changing
dipole transition, time dependent oscillations can reveal the presence of
physics beyond the Standard Model.  If $P_1$ and $P_2$ are $CP$
eigenstates ({\it e.g.} as in $B_d \to \KS \pi^0 \gamma$), then
%
%DA35(1)
%
to leading order in the effective Hamiltonian,
the oscillation is independent of the resonance structure. 
Thus data from resonances as
well as from nonresonant decays can be included.  This may significantly
enhance the sensitivity to new physics of the method.  If $P_1$ is a
charged particle, and $P_2$ its anti-particle ({\it e.g.} as in $B_d \to
\pi^+ \pi^- \gamma$), one has the additional advantage that both the
magnitude and the weak phase of any new physics contribution can be
determined from a study of the angular distribution.  These signals offer
excellent ways to detect new physics because they are suppressed in the
Standard Model. 
We also show that the potential contamination of these
signals originating from the Standard Model annihilation diagram gives
rise to photons with, to a very good approximation, the same helicity as
the dominant penguin graph and thus causes no serious difficulty.
%
%
%A+S2  
%%%%>>DA3
%
%% TJG 2004/11/07 little rewording
%ANS 36 change CP to C as that is the stronger statement
The formalism which applies to the case where $P_1$ and $P_2$ are $C$
eigenstates also further generalizes to the case 
of final states containing multiple $C$ eigenstates and a photon. 
This suggests several additional channels to search for new physics, 
%such as $\phi K_S \gamma$.
%v31
such as $K_S \eta^{\prime} (\eta) \gamma$, $\phi K_S \gamma$ {\it etc.}
%A+S2
%
%
%DA35(2)
%
%ANS36 begin
We also emphasize that the contribution of non-dipole interactions can
be monitored by the dependence of the mixing-induced $CP$ asymmetry
of non-resonant modes on the Dalitz variables.   
Furthermore, 
using a number of different final states can also provide 
important information on the contribution from non-dipole effects. 
%ANS36..end
%
%
%
\end{abstract}


\maketitle

%
%
%=================================================================
%
%

\section{Introduction}
\label{introduction}

The oscillation of neutral ground state mesons has proved to be a
sensitive probe of $CP$ violation and thus a sensitive probe for physics
at an energy scale well beyond the mass of the meson itself.  In recent
%% TJG B^0 -> B_d
years the oscillation of the neutral $B_d$ meson produced at $B$ factories
or hadronic $B$ experiments has provided a means to test the hypothesis
that the Standard Model (SM)  generates $CP$ violation~\cite{bib:sanda} 
through the
Cabibbo-Kobayashi-Maskawa (CKM) mechanism~\cite{ckm}.  In this approach
new physics (NP) becomes evident if the CKM interpretation cannot
consistently explain the results.

Within the realm of $B$ physics, radiative decays resulting from the quark
level transitions $b \to q\gamma$, where $q=d$ or $s$, have long been
recognized~\cite{bsGammaTheory} as very good channels to look for NP.  
Indeed the experimental effort~\cite{bsGammaExpt} to measure the rate of
$b \to s\gamma$ both in exclusive and inclusive channels has reached the
point where the comparison with the SM is dominated by theoretical errors.
Further reduction in the theory errors appears rather difficult. Clearly
then it is advantageous to use in addition to the rates other observables
that can reveal new physics.

One such well known observable is the direct $CP$ partial rate asymmetry.
%% TJG B -> B_d
The predicted SM asymmetry~\cite{dirCP} for $B_d \to X_s \gamma$ is rather
small, about 0.5\% and the experimental bound~\cite{bsGammaExpt} is about
%% TJG B -> B_d
a factor of ten above that. For $B_d \to X_d \gamma$ the predicted asymmetry
in the SM is actually quite large (around 15\%)~\cite{dirCP} but its
branching ratio is rather small and also it is experimentally more
challenging due to higher backgrounds.  Nonetheless, this is clearly a
promising technique in which efforts are currently being directed.

Another approach was suggested in~\cite{ags1} (AGS), and has been recently
implemented experimentally~\cite{expt_ags}, where the oscillations of the
%% TJG v21 neutral $B_d$ and $B_s$ mesons -> neutral $B$ mesons
neutral $B$ mesons are exploited.  Such oscillations in the
radiative decays of neutral $B$ mesons are suppressed in the SM so if a
significant signal of this sort is detected then NP is directly
established. Indeed the cleanliness of the NP signal in the AGS technique
provides one of the important motivations for a ``super $B$
factory''~\cite{superb_loi}.

%V27 sentence below is left from STOKES vector stuff so is being
%V27 deleted as it is now out of place 
%The key here is the fact that the photon is polarized and so in the decays
%$b\to q \gamma$ and $\OL b\to \OL q \gamma$ there are four complex
% helicity amplitudes corresponding, in principle, to seven observables.  
The key is the fact that the photon is polarized;
the short distance Hamiltonian of the SM makes the very special prediction
that photons from $b$ ($\OL b$) decay are dominantly left (right)
polarized with the same weak phase.  Thus any method such as AGS which
measures an observable that vanishes with the SM polarized photons is a null
test for new physics with little dependence on theoretical uncertainties
concerning the hadronization of the final state. Recently it was pointed
out~\cite{GH} that decays of the form $B^0 \to P^0P^0\gamma$, where $P^0$
is any spin-0 $CP$ eigenstate, can be used to probe AGS oscillations in $b
\to d\gamma$ transitions.

The problem with photons from $B$ decay is, of course, that their
polarization is not easily measured with current detector technology and
so indirect methods must be used to probe polarization dependent
observables as is the case in AGS. 
%AS25 
Besides AGS some other methods have been suggested to learn about the
photon polarization.  
Consider, for example in the case of $b
\to s\gamma$, mesonic decays of the form $B_d \to X_s\gamma$.  Because the
initial state has total angular momentum $J=0$ the helicity of the photon
must be the same as the net helicity of the $X_s$.  
In~\cite{gronauPirjolLee,GGPR} (GP) the photon polarization is probed by
considering the interferences of various $K\pi\pi$ resonances.  This
approach has an advantage that with the four body final state, parity odd
observables may be constructed but has a disadvantage that the
interpretation of the angular distributions requires some understanding of
the $1^-$, $1^+$ and $2^+$ kaonic resonances.  Another mode which may be
useful to measure the photon polarization in $b \to s\gamma$ transitions
%%v28 TJG B -> phi K gamma (not B_d ... actually B_u)
is the recently discovered $B \to \phi K \gamma$~\cite{belle_phikg}. The
photon polarization may also be studied in $\Lambda_b$
decays~\cite{HiKa,GGPR}.

Another approach suggested in~\cite{GrossmanPirjol,Sehgal} is to
``resolve'' the photon to an $e^+e^-$ pair.  This may either be
accomplished via interaction of the photon with matter through a
Bethe-Heitler conversion or internally where the photon is virtual.  
Measuring photon polarization through the Bethe-Heitler conversion may prove
experimentally challenging but would be a method of general utility.  
Furthermore, the detection of such $e^+e^-$ pairs in the vertex detector
at a $B$ factory experiment would enable determination of $B$ meson decay
positions in decay modes containing only photons in the final state.
However, since the amount of material in the inner detector must be kept
to a minimum in order to improve the experimental resolution, the rate of
such conversion is rather low. Internal conversion would be present and
dominate over short distance $e^+e^-$ production at low electron-positron
invariant mass.  The event rate would, of course, be smaller than that for
direct photons.  An additional feature of this channel would be the
ability to study the $CP$ properties of the short distance electron
positron pairs produced at a larger invariant mass~\cite{BtoKeePapers}.

In this paper, we will consider a variation on the processes considered
in~\cite{ags1} where we will assume that the final state consists of two
%% v28 TJG pseudoscalars -> pseudoscalar mesons
pseudoscalar mesons and a hard photon.  
%% v28 TJG 
For oscillations to occur, the mesons are required to be eigenstates
of charge conjugation; since they have definite parity, 
they are also $CP$ eigenstates.
The use of a hard photon offers the dual
advantage that (1)~$CP$ violation due to SM bremsstrahlung (discussed
in~\cite{Sehgal:2002ez}) is suppressed and (2)~experimental backgrounds
are suppressed at higher photon energy.  
%We also discuss how this can be
%achieved experimentally.

Of course, a final state such as $K_S\pi^0 \gamma$ is less general than the $K\pi\pi\gamma$
states considered in~\cite{gronauPirjolLee}, however, in this case a simple
analysis can lead to powerful conclusions regarding new physics.  Indeed,
%% TJG B -> B_d
in the case of $B_d \to \KS \pi^0 \gamma$, an initial study may be
%MH26
carried out as an extension of the existing analyses for
%% TJG 2004/11/07 B -> B_d
the decay $B_d \to K^{*0}\gamma,~K^{*0}\to \KS \pi^0$~\cite{expt_ags}, 
%carried out as an extension of the existing $B\to [K^{*0}\to \KS \pi^0]\gamma$ 
%analyses~\cite{expt_ags},
%MH26
%% TJG v21 try to emphasise main point
as our discussion shows that all $B_d \to \KS \pi^0 \gamma$ decays,
not only those produced via the $K^{*0}$ resonance,
may be included. 
% AS25 in the analysis.
% and our discussion has consequences for the treatment 
% of background from $B$ decay under the $K^{*0} \to \KS \pi^0$ peak. 
%ANS36 begin  
By using such a generalized final state, 
not only it is possible to increase the statistics,
but one can also extract useful information regarding the 
possible contributions to $H_{\rm eff}$
from operators other than the Standard Model dipole.
%ASv38 begin
A very important characteristic of the mixing-induced CP asymmetry
of these modes is that the contribution which 
originates from the dipole term is independent 
of the Dalitz variables.
Thus
%ASv38 end
%% TJG maybe try to say something more explicit about NP?
%ANS36 end
the key advantage of
this class of final states over other probes of polarization 
observables in $b\to q\gamma$ is that 
in this case the interpretation is relatively clean.  
In the case of $b\to s\gamma$, the SM contribution to $CP$ violating
observables is only a few percent, so a large signal would be an
unmistakable sign of new physics. For $b \to d \gamma$ the SM predicts
%% TJG v21 even much -> a much
a much 
smaller time dependent $CP$ asymmetry so that this case may be viewed as a
powerful null test.

%% TJG v21 English corrections
%AS25 chnaged as polarization Stokes vector stuff is now gone
In section II, we recall some of the basic issues
in radiative decays.
%ha systematic description of photon polarization is
%given and the constraints of $CP$ and $CPT$ on amplitudes are
%utilized.  
%% v28 TJG english correction
%%There also corrections to the dominant dipole $H_{\rm eff}$
%%and their signals are discussed.  
Corrections to the dominant dipole $H_{\rm eff}$
and their signals are also discussed.  
Section III briefly reviews the work of~\cite{ags1}, 
whose  generalization is the main focus of this paper. 
Sections IV and V contain the main body  
of our discussions on three-body modes. 
Section VI briefly mentions some generalizations and also
presents experimental considerations. 
Section VII discusses 
the helicity of photons from the annihilation contribution;
%ASv38
section ~\ref{non-dipole} briefly discusses effect of non-dipole
operators
and section IX contains a brief summary. 
The possible  complication in the analysis due to
%%the presence of a perturbative phase is relegated to a short Appendix.
the presence of a perturbative phase is very briefly
outlined in a short Appendix.

\section{Basics of radiative decays}

Let us consider radiative decays of the form $B \to F\gamma$ where $F$ is
either a single meson ({\it e.g.} $K^*$)  or a multi-particle state ({\it
e.g.} $n\pi K$).  The decay is governed by two amplitudes:  the decay to
right and left polarized photons; the same is true for the corresponding
decay of the $\OL B$.  We can denote these helicity amplitudes as follows:

%% TJG switch \gamma F \to F \gamma \ldots consistent with elsewhere, may help reader
% \begin{eqnarray}
% \OL {\cal M}_L  & = &   {\cal M}(\OL B \to \gamma_L \OL F),     \nonumber\\
% \OL {\cal M}_R  & = &   {\cal M}(\OL B \to \gamma_R \OL F), \nonumber\\
%       {\cal M}_R  & = &   {\cal M}(    B \to \gamma_R F),     \nonumber\\
%       {\cal M}_L  & = &   {\cal M}(    B \to \gamma_L F).
% \label{genericAmplitudes}
% \end{eqnarray}
\begin{eqnarray}
\OL {\cal M}_L  & = &   {\cal M}(\OL B \to \OL F \gamma_L), \nonumber\\
\OL {\cal M}_R  & = &   {\cal M}(\OL B \to \OL F \gamma_R), \nonumber\\
    {\cal M}_R  & = &   {\cal M}(    B \to     F \gamma_R), \nonumber\\
    {\cal M}_L  & = &   {\cal M}(    B \to     F \gamma_L).
\label{genericAmplitudes}
\end{eqnarray}

\noindent Here we adopt a phase convention for $\gamma_{L,R}$ such that
their phases are equal if the parity of the final state is opposite to the
internal parity of $F$.

At short distances, the photons arise from a radiative transition of the
$b$ quark.  In most models for this process arising from the electroweak
scale or higher, it is to be expected that the dominant contribution to
%% TJG v21 5 \to five (use word)
the $b$-scale effective Hamiltonian is via the dimension five dipole
transition operator,
%
%DA35(3)
%
which we will initially assume dominates the process:

\begin{equation}
  H_{\rm eff} = - \sqrt{8} G_F \frac{e m_b}{16\pi^2} F_{\mu\nu}
  \left [
    F_L^q\ \OL q\sigma^{\mu\nu}\frac{1+\gamma_5}{2}b
    +
    F_R^q\ \OL q\sigma^{\mu\nu}\frac{1-\gamma_5}{2}b
  \right ] + h.c.
  \label{effective_H}
\end{equation}

\noindent Here $F_L^q$ ($F_R^q$) corresponds to the amplitude for the
emission of left (right) handed photons in the $b_R \to q_L \gamma_L$
($b_L \to q_R \gamma_R$) decay, {\it i.e.} in the $\OL B \to \OL F
\gamma_L$ ($\OL B \to \OL F \gamma_R$) decay. 
%% v28 TJG make clear Eq. 3 is definition of psi^q
%We can write $F_L^q$ and
%$F_R^q$ in terms of a parameter $\psi^q$, 
We can relate $F_L^q$ and $F_R^q$ by defining a parameter $\psi^q$,
which is ${\cal O}\left(\frac{m_q}{m_b}\right)$ in the SM:

\begin{eqnarray}
  F_L^q & = & F^q e^{i\phi_L^q} \cos \psi^q, \nonumber\\
  F_R^q & = & F^q e^{i\phi_R^q} \sin \psi^q,
  \label{psidef}
\end{eqnarray}

\noindent 
where $\phi^q_L$ and $\phi^q_R$ are $CP$ violating phases. 
Since the strong interaction respects parity and charge conjugation, 
this model implies that the amplitudes are given in terms of a single 
%% v28 TJG english
complex valued form factor $f$, so that:

%AS25 important changes in eq. below 
\begin{eqnarray}
  \OL {\cal M}_L & = & - F_L^{q}  f(P\Phi_F)[P_{\rm internal}], \nonumber\\
  \OL {\cal M}_R & = &   F_R^{q}  f(\Phi_F), \nonumber\\
    {\cal M}_R & = & F_L^{q*} f(  C\Phi_F)[C_{\rm internal}], \nonumber\\
      {\cal M}_L & = & -F_R^{q*} f(CP \Phi_F)[(CP)_{\rm internal}].
    \label{MShortDistance}
\end{eqnarray}

%AS25 added para below
%MH26 definition of $\Phi_F$ restored.
%V27 definition of $\Phi_F$ is sharpened
\noindent Here $\Phi_F$ represents 
the phase space of 
%a set of observables that determine
the final state $F$.
%MH26
The sign in front of $F_L^{q}$ and $F_R^{q*}$
arises from the negative
parity of the initial $B^0$ or $\bar B^0$ state.
%V27..delete phrase below (its too late to put it here) and replace with
%V27 Also  
%In the above equations 
Also, $P_{\rm internal}$, $C_{\rm internal}$,
$(CP)_{\rm internal}$ are the internal $P$, $C$, $CP$ eigenvalues of all final
state particles which are $P$, $C$, $CP$ eigenstates.
%%\noindent where $\OL {\cal M}_L$ is the amplitude for the decay of
%% TJG v21 English & B^0 \to B
%%a $\OL B$ into the $F \gamma_L$ final state, {\it etc.}      
%AS25 delete sentence below 
%The universality of the form factor above is a prediction which
%should be tested experimentally.  
%MH26 add examples of C_internal and P_internal
For example, in case of 
$B_d \to \KS \pi^0 \gamma_L$ and
$B_d \to \KS \pi^0 \gamma_R$ decays, 
we obtain 
\begin{eqnarray}
  P_{\rm internal} &=& P_{\KS}P_{\pi^0} = (-1)(-1) = +1,\nonumber \\
  C_{\rm internal} &=& C_{\KS}C_{\pi^0}C_{\gamma} = (-1)(+1)(-1) = +1,
\end{eqnarray}
%% v28 tjg english
where we use $J^{PC} = 0^{--}$ for $\KS$ ignoring the tiny 
$CP$ violation effect in the neutral kaon system.
%MH26
The strong phases in $f$ arise from
the rescattering between
mesons in the final state $F$.

\begin{figure}
  \epsfxsize 2.9 in
  \mbox{\epsfbox{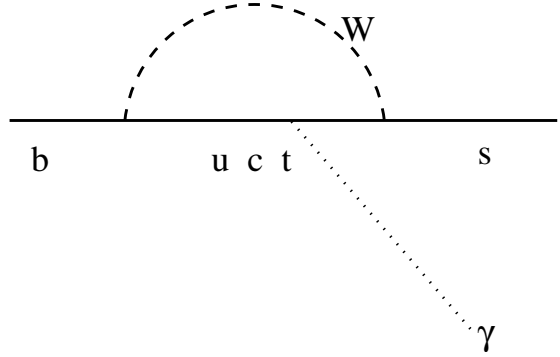}}
  \caption{
    A typical SM radiative penguin graph.
  }\label{feydia_a}
\end{figure}

In the SM, the contributions are predominantly given by penguin diagrams
%%V30 MH Unnecessary () removed from Fig.~(\ref{feydia_a}); done also for
%%V30 all other places.
such as that shown in Fig.~\ref{feydia_a}.  By CKM unitarity the short
distance contribution to this has a CKM phase given by the phase of
$V_{tb}V_{ts}^*$.  This short distance contribution yields predominantly
left-handed photons with the right handed component suppressed by
$m_s/m_b$. This right-handed component will also have the same 
weak phase as
the left-handed component.

\begin{figure}
  \epsfxsize 2.9 in
%% TJG v21 try to make cut line prettier \ldots not very successful
%   \mbox{\epsfbox{ags2_fd_b.eps}}
  \mbox{\epsfbox{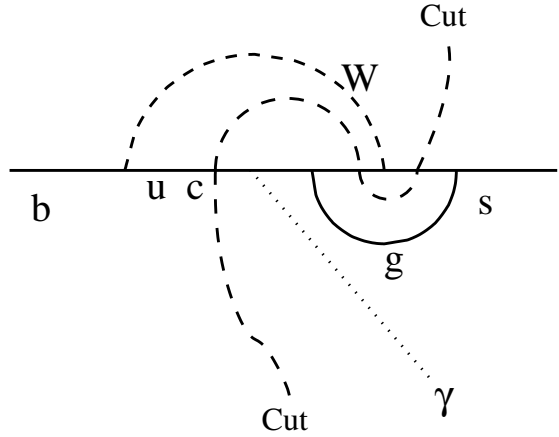}}
  \caption{
    An example of a QCD loop correction which generates the absorptive
part necessary for direct $CP$ asymmetry, see~\cite{dirCP}. The cut is
indicated by the dashed line.
  }\label{feydia_b}
\end{figure}

\begin{figure}
  \epsfxsize 2.9 in
  \mbox{\epsfbox{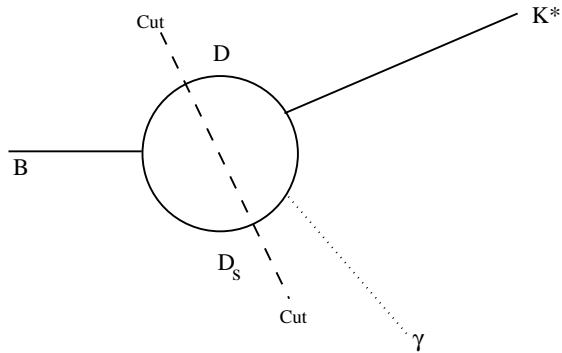}}
  \caption{
    A possible long distance rescattering effect due to on-shell $D_s\OL D$
contribution, see reference~\cite{grinsteinPirjol}.
  }\label{feydia_c}
\end{figure}

\begin{figure}
  \epsfxsize 2.9 in
  \mbox{\epsfbox{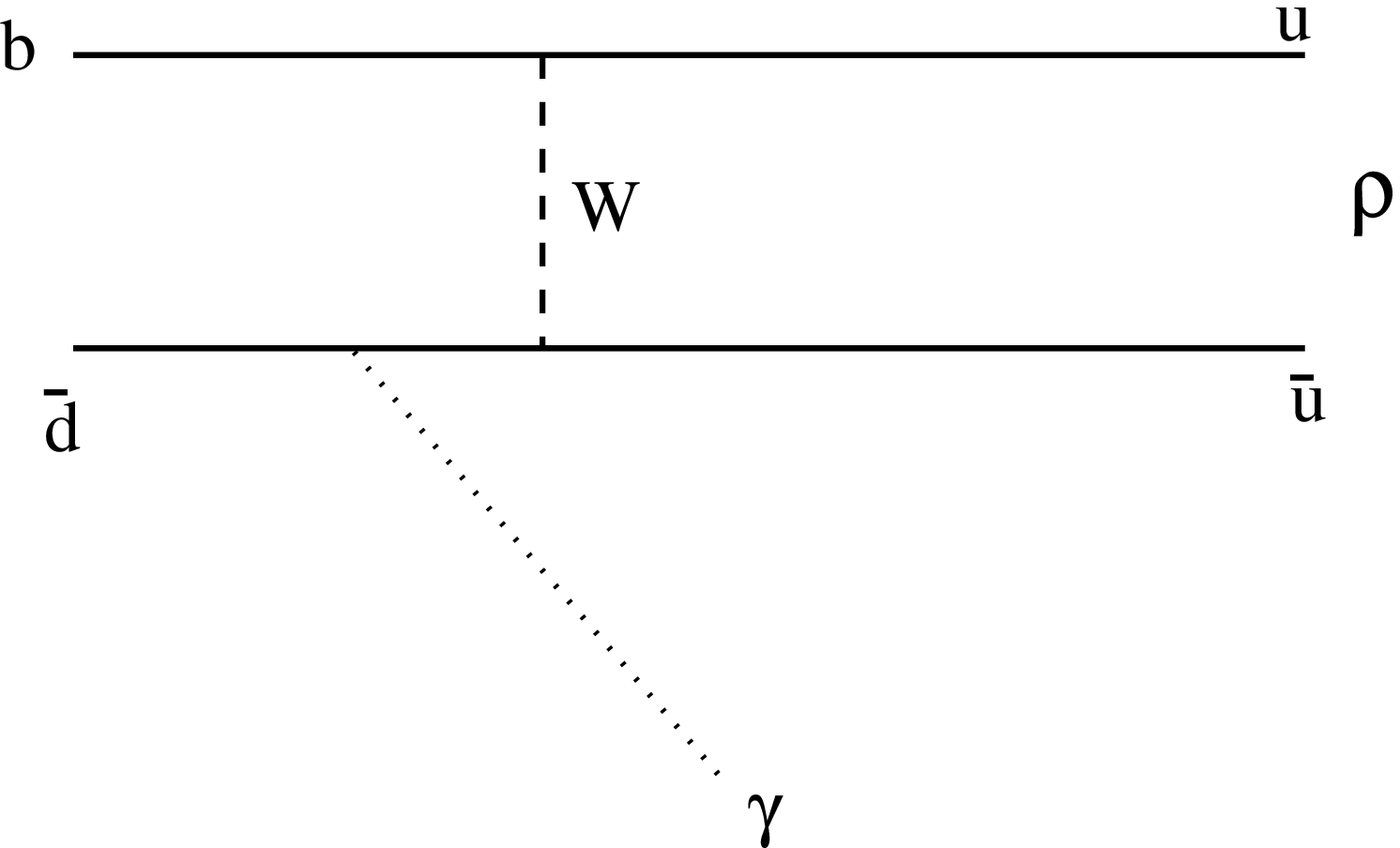}}
  \caption{
    The dominant contribution from the annihilation graph, see
reference~\cite{bloc}.
  }\label{feydia_d}
\end{figure}

The long distance (LD) contributions arise from the $c$ and $u$ penguins.  
They can have a nontrivial rescattering phase distinct from those present
in the form factor $f$ mentioned above.  An example of a quark diagram
contributing is shown in Fig.~\ref{feydia_b}, which can also be
understood as a rescattering of mesons through processes such as that
shown in Fig.~\ref{feydia_c}.
%% v28 TJG remove indeed 
%%Indeed Fig.~\ref{feydia_d}
Fig.~\ref{feydia_d}
represents another LD contribution to radiative  
decays arising from ``annihilation" diagrams; we shall 
discuss this particular contribution later in the paper. 

%
%AS25 wording change
%% v28 tjg ``such processes'' -> such LD processes
The precise effect of such LD processes is difficult to calculate reliably.
In~\cite{grinsteinPirjol} a detailed estimate in the case of $B\to
%% v28 TJG photon -> photons
V\gamma$ is given and it is found that the photons from these
%% v28 TJG remove LD
contributions are still predominantly left-handed. On the other hand, the
authors of~\cite{Sehgal} entertain the possibility that the contribution
is of the form ${\cal M}_L = -{\cal M}_R$.

In the case of $b \to s\gamma$ such a contribution will be
suppressed with respect to the short distance~\cite{GreubSimmaWyler}.  
However, for
$b \to d\gamma$ the magnitude of such a contribution may be
%%v30 MH quite appreciable.  Phenomenologicaly, the primary manifestation 
quite appreciable.  Phenomenologicaly, the primary manifestation 
of such ``long
distance" contributions would be partial rate asymmetries in $B \to
K^* \gamma$ or $B \to \rho\gamma$.

In such cases when there is a contribution which has a strong phase, it can
result in a contribution to
%% TJG v30
Eqn.~\ref{MShortDistance} which although 
cannot readily be reliably calculated, may be parameterized as follows:

\begin{eqnarray}
  \OL {\cal M}_L & = & -F_L^{q } f(P\Phi_F) 
  (1 + \OL \delta_L(P\Phi_F))[P_{\rm internal}],
  \nonumber\\
  \OL {\cal M}_R & = & F_R^{q } f(\Phi_F) (1 + \OL \delta_R(\Phi_F)),
  \nonumber\\
      {\cal M}_R & = & F_L^{q*} f(C\Phi_F) 
      (1 +     \delta_R(C \Phi_F))[C_{\rm internal}],
  \nonumber\\
      {\cal M}_L & = & -F_R^{q*} f(CP\Phi_F) 
      (1 +     \delta_L(CP \Phi_F))[(CP)_{\rm internal}],
  \label{MGeneral}
\end{eqnarray}

\noindent where the $\delta$ terms are arbitrary complex form factors.  
If such contributions are only from the SM and the photon is predominantly
the same as the short distance SM effects then:

\begin{eqnarray}
      \delta_L & = & \OL \delta_R = 0,   \nonumber\\
      \delta_R & = & e^{ i\mu} \Delta,     \nonumber\\
  \OL \delta_L & = & e^{-i\mu} \Delta,
\end{eqnarray}

\noindent where $\mu$ is the weak phase between the charm and top penguins
and $\Delta$ is a complex valued function of $\Phi_F$.
%AS25..delete sentence left from STOKE%
%In the case of $b \to s\gamma$ of the seven possible polarization
%observables there is one degree of freedom dominant in the SM so any
%significant deviation from this situation is a sure indication of NP.  
%AS25 clean up language below  
For $b \to s\gamma$ the small direct
partial rate asymmetry from the SM 
may also be eventually
detectable~\cite{dirCP,Sehgal:2002ez}.  
In the case of $b\to d\gamma$ the 
%% v28 TJG CP -> $CP$
direct $CP$ asymmetry can be quite sizable. 

In general, new physics should assert itself as an additional contribution
%% TJG v30
to $F^q$ and $\phi$ in Eqn.~\ref{psidef} which is different from
the SM and we will assume therefore that new physics is only manifest at
short distances.  If this were not the case, then probably other signals
would be more suited for its detection.  Since our methods are
generalizations of those proposed in AGS~\cite{ags1}, let us now briefly
review that method.

\section{AGS Oscillation}
%% MH v21:
% B -> M gamma is replaced with B -> V gamma to avoid confusion
% between this M meson and M for mixing.

Before proceeding it is useful to consider the conditions under which the
AGS oscillations occur.  These follow from the general conditions for
oscillating signals in neutral mesons applied specifically to the case of
radiative decays.  Thus for there to be oscillations in the decay $B \to
V\gamma$ the following conditions are necessary:

%AS25 expand bracket below
%% TJG v21 K0* -> K*0
\begin{enumerate}
\item[(1)] 
  Both $B$ and $\OL B$ must decay to the same 
  exclusive final state $V\gamma$ ({\it e.g.} 
%V31 remove second half of sentence as it is repeated several lines
%below
  $V = K^*$, $\rho$ or higher resonances).   
 % with the same
  %quantum content); 
% further note 
%% v28 TJG K^{0*} -> K^{*0}
%that in the case of $B\to K^{*0} \gamma$, the $K^{*0}$
%must decay to a $CP$ eigenstate, for instance $ K^{*0}\to \KS\pi^0$.
\item[(2)] 
  The photons produced in $B \to V\gamma$ or $\OL B \to V\gamma$ 
  must be a mixture of right and left handed helicities.
\end{enumerate}

%
%
%============----->>>>>>>here<<<<<<<-------------
%
%

Assuming these conditions are met, we recall that in general we can
describe the time dependent wave function of the $B$ meson (either $B_d$
or $B_s$) as~\cite{pdg_review}:

\begin{eqnarray}
  |    B(t) \ket & = & g_+ |    B \ket + \frac{q}{p}g_- |\OL B \ket,
  \nonumber\\
  |\OL B(t) \ket & = & g_+ |\OL B \ket + \frac{p}{q}g_- |    B \ket,
\end{eqnarray}

where

\begin{eqnarray}
  g_\pm = \frac12 e^{-i M_1 t} e^{- \frac{1}{2} \Gamma_1 t} 
  \left [
    1 \pm e^{-i \Delta m t}e^{\frac{1}{2} \Delta \Gamma t}
  \right ].
\end{eqnarray}

%% v28 TJG tidy up
%In the case of a $B_d$, we will assume that $\left| q/p \right| = 1$,
%%AS25 add line below
%that $q/p = e^{i \phi_M}$, with $\phi_M = - 2 \phi_1$~\cite{abc}
%in the SM, 
%and that $\dgd$ is small. In this limit we can write the time dependent decay
%rates of $B_d$ and $\OL B_d$ to given final state $V\gamma$ as:
In the case of $B_d$ decay,
we assume that $q/p = e^{i \phi_M}$, with $\phi_M = - 2 \phi_1$~\cite{abc}
in the SM,  and that $\dgd$ is small. 
In this limit we can write the time dependent decay
rates of $B_d$ and $\OL B_d$ to given final state $V\gamma$ as:

%{\bf TJG v21: 
%  I think we should give explicit rates for $\gamma_L$ and $\gamma_R$ here,
%  then add them.}
%% MH v22: M_R and M_L are explicitly shown.

\begin{eqnarray}
  \Gamma_{B_d \to V\gamma}(t)  \equiv  \Gamma_{B_d \to V\gamma_L} (t) 
                     + \Gamma_{B_d \to V\gamma_R} (t)
            & \propto &
  e^{-\Gamma t}
  \left[ X_{V\gamma} + Y_{V\gamma}\cos(\dmd t) + Z_{V\gamma}\sin(\dmd t) \right],
  \nonumber\\
  \Gamma_{\OL B_d \to V\gamma}(t)  \equiv  \Gamma_{\OL B_d \to V\gamma_L} (t) 
                         + \Gamma_{\OL B_d \to V\gamma_R} (t)
                & \propto & 
  e^{-\Gamma t}
  \left[ X_{V\gamma} - Y_{V\gamma}\cos(\dmd t) - Z_{V\gamma}\sin(\dmd t) \right],
\label{eq:decayrate}
\end{eqnarray}

where
\begin{eqnarray}
  X_{V\gamma} & = & (|{\cal M}_L|^2 + |{\cal M}_R|^2) 
                  + (|\OL{\cal M}_L|^2 + |\OL{\cal M}_R|^2),
  \nonumber\\
  Y_{V\gamma} & = & (|{\cal M}_L|^2 + |{\cal M}_R|^2)
                  - (|\OL{\cal M}_L|^2 + |\OL{\cal M}_R|^2),
  \nonumber\\
  Z_{V\gamma} & = & -2 \, \Im \left( e^{i\phi_M}
                                    ( {\cal M}_L^* \OL{\cal M}_L
                                     +{\cal M}_R^* \OL{\cal M}_R) \right).
\end{eqnarray}

%% MH v22: Modified
%\noindent Here ${\cal M}$ is the amplitude for $B_d \to V\gamma$ and
%$\OL{\cal M}$ is the amplitude for $\OL B_d \to V\gamma$.
\noindent Here we sum decay rates for the left-handed and
right-handed photon helicity states as
we do not distinguish between the two.
%MH26 Definition of {\cal S} and {\cal A} 
%% TJG 2004/11/07 $CP$-violation -> $CP$ violation
We also define $CP$ violation parameters ${\cal S}_{V\gamma}$
and ${\cal A}_{V\gamma}$ as~\cite{asp_def}
%V27....change = to \equiv
\begin{equation}
\frac{\Gamma_{\OL B_d \to V\gamma}(t) - \Gamma_{B_d \to V\gamma}(t)}
{\Gamma_{\OL B_d \to V\gamma}(t) + \Gamma_{B_d \to V\gamma}(t)} \equiv 
  {\cal S}_{V\gamma}\sin(\dmd t) + {\cal A}_{V\gamma}\cos(\dmd t).
  \label{calsdef}
\end{equation}

\noindent The parameter ${\cal S}_{V\gamma}$ represents mixing-induced
$CP$ violation, while ${\cal A}_{V\gamma}$ represents direct
$CP$ violation.
%% TJG v30
From Eqn.~\ref{eq:decayrate}, we obtain
${\cal S}_{V\gamma} = -Z_{V\gamma}/X_{V\gamma}$ and
${\cal A}_{V\gamma} = -Y_{V\gamma}/X_{V\gamma}$.
%MH26

%% v28 TJG english
In AGS it was assumed that $V\gamma$ was a vector-photon ($K^*\gamma$
or $\rho\gamma$)  final state and that the photon emission was described
%% TJG v30
by the short distance Hamiltonian of Eqn.~\ref{effective_H}.  Note that
to respect the conditions for oscillations, the $K^*$ state must be in a
%AS25 add footnote for K_L^* state
%V30 CP -> below
$C$ eigenstate which is only true if $K^*\to \KS\pi^0$ or $K^*\to
\KL\pi^0$ (the latter is unlikely to be 
%MH26 KL_state footnote put into the text.
%experimentally useful)~\cite{KL_state}. 
experimentally useful). 
%% v28 TJG remove $K_S^*$ and $K_L^*$ - we don't need this notation
%The $J^{PC}$ of $K^*$ when it decays to $\pi^0 K_s$ and $\pi^0 K_L$ are
%$1^{--}$ and $1^{-+}$, respectively; we denote these as $K_S^*$ and $K_L^*$.  
%MH26
Assuming
that the polarization information 
%discussed
%in~\cite{GrossmanPirjol,Sehgal} 
is not available, we must content
ourselves with summing over polarization and, neglecting long distance
contributions, thus obtain the time dependent forms:

%% TJG v21 include q superscripts and \chi_V to indicate CP of V
%AS25 switch signs of x_i below
\begin{eqnarray}
  \Gamma_{B_d \to V\gamma}(t) & \propto & e^{-\Gamma t} 
  \left[ 1 + \chi_V \sin 2\psi^q \sin \phi^q \sin(\dmd t) \right],
  \nonumber \\
  \Gamma_{\OL B_d \to V\gamma}(t) & \propto & e^{-\Gamma t} 
  \left[ 1 - \chi_V \sin 2\psi^q \sin \phi^q \sin(\dmd t) \right],
\label{xi}
\end{eqnarray} 

\noindent
%%MH v21: Footnote on the definition of the phi_M is added.
%V27...superscript q is needed on LHS below.  
where $\phi^q=\phi_M+\phi_L+\phi_R$~\cite{phase_def}, 
%AS25 C not CP
%% TJG 2004/11/07 add footnote 
$\chi_V$ denotes the $C$ eigenvalue of $V$~\cite{c_not_cp}
%% TJG v21 include q superscripts
and the superscript $q$ indicates the quark produced in the $b \to q\gamma$ decay.

This oscillation therefore allows the extraction of the quantity
%% TJG v21 add S=Z/X
$\cals_{V\gamma} = -\chi_V \sin 2\psi^q \sin \phi^q$.
%MH26 $\cals_{V\gamma} = \chi_V \sin 2\psi^q \sin \phi^q$, which is related to the 
%MH26 quantities used above by $\cals_{V\gamma} = -Z_{V\gamma}/X_{V\gamma}$. 
In the case of $B_d \to K^*\gamma$
where the short distance contribution to the photon is predominantly right
%% TJG v21 removed ``therefore'' which is in preceeding and following lines
handed, $\psi^s$ is small and so $\cals_{K^*\gamma}$ is consequently
small.  The observation of a significant $\cals_{K^*\gamma}$ would therefore
indicate the presence of NP.  To the extent that the long distance
contribution to the photon is right handed as suggested by the calculation
of~\cite{grinsteinPirjol}, then the same is also true of $B_d \to \rho\gamma$.

Let us now consider the generalization to final states with two
pseudoscalars, in particular $\KS \pi^0 \gamma$ and $\pi^+\pi^- \gamma$.
%% v28 TJG removed K_S^* notation
Note that the case of $K^*\gamma$ with $K^* \to K_S \pi^0$
is a special case of $K_S \pi^0 \gamma$
where the two mesons are on the $K^*$ resonance while $\rho^0\gamma$ is a
special case of $\pi^+\pi^- \gamma$ with the two mesons on the $\rho$
resonance.
%AS25 add new significance of xi 
For clarity, let us remark that any kaonic
%% v28 TJG typo fixed
resonance of angular momentum $J$, 
that produces $K_S \pi^0$, will have $P = (-1)^J$ with
$C = -1$; thus for any such resonance $\chi_V = -1$ and so 
the oscillations for all of them will
%% TJG v30
be identical as they are all governed by Eqn.~\ref{xi}.  

In contrast, for a $\pi^+ \pi^-$ resonance of spin $J$,
$P =(-1)^J$ and (since $CP = + $) $C = (-1)^J$. It follows
then that $\chi_V = (-1)^J$, so all odd-$J$ and even-$J$ resonances
%% TJG v30
will have opposite signs in Eqn.~\ref{xi}.

In the next section we will generalize these results to be
independent of the resonance the mesons might go through
and consider the angular distribution
of the pseudoscalars.   
%AS25 insertion on xi end

%% TJG B^0 -> B_d
\section{\boldmath $B_d \to \KS\pi^0\gamma$ and $B_d \to \pi^+\pi^-\gamma$}
\label{sectionByPP}

In this section we will contrast the nature of AGS oscillations in the
%% TJG B^0 -> B_d
%AS25, add footnote here, pi^0 can always be replaced by eta,eta'
cases of $B_d \to \KS\pi^0 \gamma$~\cite{also_eta}
and $B_d \to \pi^+\pi^- \gamma$. 
The discussion here is easily generalized to the case of other 
(pseudo-)scalar pairs recoiling against the photon.  
We will consider the more general case including direct $CP$ violation
in the next section.

The key point to realize is that there is a contrast between the symmetry
properties of phase space in the case of $\KS\pi^0$ with that from
$\pi^+\pi^-$.  To see how this arises let us designate the $\KS$ to be
particle ``1'' and the $\pi^0$ to be particle ``2'' in the first case
while we designate $\pi^+$ as particle ``1'' and $\pi^-$ as particle ``2''
in the second case.  For $\KS\pi^0 \gamma$ then:

\begin{eqnarray}
  C\Phi(\gamma_{^R_L},1,2) & = & \Phi(\gamma_{^R_L},1,2), \nonumber\\
  P\Phi(\gamma_{^R_L},1,2) & = & \Phi(\gamma_{^L_R},1,2),
\end{eqnarray}

while for $\pi^+\pi^- \gamma$, 

\begin{eqnarray}
  C\Phi(\gamma_{^R_L},1,2) & = & \Phi(\gamma_{^R_L},2,1), \nonumber\\
  P\Phi(\gamma_{^R_L},1,2) & = & \Phi(\gamma_{^L_R},1,2).
\end{eqnarray}

%%TJG v21 add emphasis on this point
\noindent
%That is to say, under the $CP$ transformation,
%each point in the $\KS\pi^0 \gamma$ phase space is 
%translated into the same point
%(notwithstanding the distinct photon helicity),
%whereas for $\pi^+\pi^-\gamma$ the positions of the pions are interchanged.

%V29 change last para to as follows

That is to say, under the $C$ and $P$ transformations,
each point in the $\KS\pi^0 \gamma$ phase space is translated into the
same point
(notwithstanding the distinct photon helicity under $P$ and $CP$),
whereas for $\pi^+\pi^-\gamma$ the positions of the pions are
interchanged under $C$ and $CP$.

%% TJG v21 ``generalize this'' -> ``generalize the AGS formalism''
It is now a simple matter to generalize the AGS formalism
to the case where the final amplitude is also a function of phase space.  
%% TJG v21 I think this is repetitive and not necessary
% Some care must be taken
% however since the decay $B_d \to \KS\pi^0 \gamma$ has different behavior
% than $B_d \to \pi^+\pi^- \gamma$.
%% TJG v21 remove paragraph break.
Consider first the case of $B_d \to \KS\pi^0 \gamma$, which at the quark
level corresponds to $\OL b \to \OL s \gamma$.  The decay amplitude
consists of two components which do not interfere with each other
%% TJG v21 English
corresponding to photons with left and right handed helicities.  
These amplitudes
should depend on the Dalitz plot variables which we will denote:

\begin{eqnarray}
  s_1 & = & (p_{\KS} + p_{\pi_0})^2, \nonumber\\
  s_2 & = & (p_{\KS} + p_{\gamma}  )^2, \nonumber\\
  s_3 & = & (p_{\gamma}   + p_{\pi_0})^2, \nonumber\\
  z   & = & \frac{s_3-s_2}{s_3+s_2}.
  \label{dalitz1}
\end{eqnarray}

\noindent
In particular the amplitude can be expressed as a function of $s_1$ and $z$, 
%% v28 TJG say explicitly what s_1 is, since also do for z
where $s_1$ is the invariant mass squared of the $\KS\pi^0$ system,
and $z$ is the cosine of the angle between the $B_d$ and $\pi^0$ in the 
$\KS\pi^0$ frame.

If we assume that the decay $B_d\to \KS\pi^0\gamma$ is governed by
%% TJG v30
Eqn.~\ref{effective_H} then the amplitude as a function of $s_1$ and $z$
%V27 footnote added reg. ref. frame 
%% TJG v30
can be written from Eqn.~\ref{MShortDistance} as~\cite{ref_frame1}:
%MH26 can be written as:
%V27...f_L and f_R need to be switched in two eqns below 

\begin{eqnarray}
  {\cal M}_R(s_1,z) & = & F^{s*}_L f_R(s_1,z), \nonumber\\
  {\cal M}_L(s_1,z) & = & F^{s*}_R f_L(s_1,z), 
\end{eqnarray}

%% TJG 2004/11/07 specify subscript on ${\cal M}$ 
\noindent where the subscript on ${\cal M}$ 
indicates the helicity of the photon emitted
and the superscript indicates the quark produced by the reaction 
({\it i.e.} $d$ or $s$). 
%MH26 QCD respects parity and the structure of $f_L$ and $f_R$ depends only on QCD, 
%MH26 therefore $f_L(s_1,z) = f_R(s_1,z) = f(s_1,z)$. 
%MH26
For the charge conjugate decay  $\OL B_d \to \KS\pi^0 \gamma$ we can likewise 
write:
 
\begin{eqnarray}
  \OL {\cal M}_R(s_1,z) & = & F^{s }_R \OL f_R(s_1,z), \nonumber\\
  \OL {\cal M}_L(s_1,z) & = & F^{s }_L \OL f_L(s_1,z). 
\end{eqnarray}

\noindent Since QCD respects both $C$ and $P$, we expect from 
Eqn.~\ref{MShortDistance}
%MH26 \noindent Again, since QCD respects $C$, $P$ we expect that 
%% TJG v21 CP(\KS) = \KS \ldots use \xi to denote the eigenvalues
%%$\OL f_L(s_1,z) = \OL f_R(s_1,z) = CP(\pi^0)CP(\KS)f(s_1,z) = -f(s_1,z)$.  
%AS25 changed signs below

\begin{equation}
f_R(s_1,z) = \OL f_R(s_1,z) = -f_L(s_1,z) = - \OL f_L(s_1,z).
\end{equation}

\noindent Thus we obtain

%AS25 xi's not relevant any more
%where $\xi_{\pi^0}$ and $\xi_{\KS}$ denote 
%the $CP$ eigenvalues of $\pi^0$ ($-1$) and $\KS$ ($+1$, for our purposes),
%respectively.  
%MH26 Moved downward.
%MH26 Note that in this
%MH26 discussion of the amplitudes at a fixed point $(s_1,z)$ in phase space, the
%MH26 relative angular momentum between $\KS$ and $\pi^0$ does not
%% TJG v21 remove ``explicitly'' \ldots neither does it enter implicitly (?)
%% explicitly
%MH26 enter and therefore the quantum
%MH26 numbers ({\it e.g.} spin) of intermediate kaonic resonances
%MH26 contributing to the process do not 
%MH26 effect the conclusions~\cite{ags_diff}.  In summary,
%MH26 these amplitudes are therefore:

%%MH v22: write xi_ks and xi_pi0 explicitly.
%\begin{eqnarray}
%  \OL {\cal M}_L(s_1,z) & = & -F^{s }_L f(s_1,z), \nonumber\\
%  \OL {\cal M}_R(s_1,z) & = & -F^{s }_R f(s_1,z), \nonumber\\
%AS25 removed xi and fix signs
%\OL {\cal M}_L(s_1,z) & = &\xi_{\KS}\xi_{\pi^0} F^{s }_L 
%f(s_1,z), \nonumber\\
% \OL {\cal M}_R(s_1,z) & = &\xi_{\KS}\xi_{\pi^0} F^{s }_R 
%f(s_1,z), \nonumber\\

%   \OL {\cal M}_L(s_1,z) & = &- F^{s }_L f(s_1,z), \nonumber\\
%    \OL {\cal M}_R(s_1,z) & = &F^{s }_R f(s_1,z), \nonumber\\

%  \OL {\cal M}_R(s_1,z)     & = &  F^{s*}_L f(s_1,z), \nonumber\\
%  \OL {\cal M}_L(s_1,z)     & = &-  F^{s*}_R f(s_1,z). \nonumber\\ 
%\end{eqnarray}

\begin{eqnarray}
  \OL {\cal M}_L(s_1,z) & = & -F^{s }_L f(s_1,z), \nonumber\\
  \OL {\cal M}_R(s_1,z) & = & F^{s }_R f(s_1,z), \nonumber\\
      {\cal M}_R(s_1,z)     & = &  F^{s*}_L f(s_1,z),\nonumber\\
      {\cal M}_L(s_1,z)     & = &-  F^{s*}_R f(s_1,z),
\end{eqnarray}

\noindent where we define a universal form factor $f(s_1,z) = f_R(s_1,z)$.
Note that in this
discussion of the amplitudes at a fixed point $(s_1,z)$ in phase space, the
relative angular momentum between $\KS$ and $\pi^0$ does not
enter and therefore the quantum
numbers ({\it e.g.} spin) of intermediate kaonic resonances
contributing to the process do not 
effect the conclusions~\cite{ags_diff}.  
%MH26 Removed to avoid redundancy.
%MH26 In summary,
%MH26 these amplitudes are therefore:
%MH26 Note again that the orbital angular momentum does not enter these expressions 
%MH26 since it is true for each point in phase space. 
%This contradicts the
%onclusion of \cite{ags1} and 
This is somewhat different from the behavior, discussed
below, for the case where the two final state mesons are 
antiparticles {\it e.g.}
$B_d\to \pi^+\pi^- \gamma$.

The time dependent rates for physical $B_d$ and $\OL B_d$ decays,
at a point in phase space defined by $\left( s_1, z \right)$, 
and summed over photon helicity are therefore given by:

%AS25 at fixed s1, z needs to be displayed out
%% TJG B^0 -> B_d
%% v28 TJG remove \Gamma(t,s_1,z)
\begin{eqnarray}
  \Gamma_{    B_d \to \KS \pi^0 \gamma}(t,s_1,z) & \propto & 
  e^{-\Gamma t}
  \left[ 
    X_{\KS \pi^0 \gamma}(s_1,z) + Y_{\KS \pi^0 \gamma}(s_1,z)\cos(\dmd t) + Z_{\KS \pi^0 \gamma}(s_1,z)\sin(\dmd t) 
  \right],
  \nonumber\\
  \Gamma_{\OL B_d \to \KS \pi^0 \gamma}(t,s_1,z) & \propto & 
  e^{-\Gamma t} 
  \left[
    X_{\KS \pi^0 \gamma}(s_1,z) - Y_{\KS \pi^0 \gamma}(s_1,z)\cos(\dmd t) - Z_{\KS \pi^0 \gamma}(s_1,z)\sin(\dmd t) 
  \right],
  \label{genericA}
\end{eqnarray}

\noindent
%% TJG, again, make explicit that LD is neglected
where, neglecting long distance effects,

%% MH v12: Z = +2(F^s)^2.... based on the modified definitions
%\begin{eqnarray}
%  X_{\KS \pi^0 \gamma} & = & 2(F^s)^2 |f(s_1,z)|^2,  \nonumber\\
%  Y_{\KS \pi^0 \gamma} & = & 0,                      \nonumber\\
%  Z_{\KS \pi^0 \gamma} & = & - 2(F^s)^2 |f(s_1,z)|^2 \sin 2\psi^s \sin \phi^s,
%\end{eqnarray}

%% TJG v21 explicitly add CP of \KS and \pi^0
%AS25 xi's for ks pi0 are superflous so are being removed 
%AS25 s_1,z dependence 

\begin{eqnarray}
  X_{\KS \pi^0 \gamma}(s_1,z) & = & 2(F^s)^2 |f(s_1,z)|^2,  \nonumber\\
  Y_{\KS \pi^0 \gamma}(s_1,z) & = & 0,                      \nonumber\\
  Z_{\KS \pi^0 \gamma}(s_1,z) & = & 
  - 2(F^s)^2 |f(s_1,z)|^2 \sin 2\psi^s \sin \phi^s,
%Z_{\KS \pi^0 \gamma} & = &
%  - 2(F^s)^2 \xi_{\KS} \xi_{\pi^0} |f(s_1,z)|^2 \sin 2\psi^s \sin
%  \phi^s,
\end{eqnarray}

\noindent
and $\phi^s = \phi_M + \phi_L^s + \phi_R^s$ is the weak phase. 

Thus, for each value of $s_1$ and $z$, 
%% TJG remove independent of photon helicity, which seems confusing to me
%% the $CP$ asymmetry~\cite{asp_def} independent of photon helicity is:
the $CP$ asymmetry is:
%MH26 the $CP$ asymmetry~\cite{asp_def} is:

%% TJG remove A(t) which is not used again
%AS25 correct sign below
%AS25 also again display s_1,z dependence and lack therof
%% v28 TJG replace  \Gamma(t,s_1,z) with \Gamma_{\OL B_d \to \KS \pi^0 \gamma}
\begin{equation}
%%   A(t) = 
  \frac{
    \Gamma_{\OL B_d \to \KS \pi^0 \gamma} - \Gamma_{B_d \to \KS \pi^0 \gamma}
  }{
    \Gamma_{\OL B_d \to \KS \pi^0 \gamma} + \Gamma_{B_d \to \KS \pi^0 \gamma}
  } = 
%%MH v22: minus sign needed  -\sin 2\psi^s \sin \phi^s \sin(\dmd t).
  + \sin 2\psi^s \sin \phi^s \sin(\dmd t).
  \label{aeqn}
\end{equation}

\noindent Note that this expression is true whether the $\KS\pi^0$ 
%% TJG remove ``continuum''
is produced by the decay of a resonance or is nonresonant.  
Furthermore, the fact that the effective Hamiltonian 
%% TJG v30
of Eqn.~\ref{effective_H} produces the photons implies that this 
asymmetry does not depend on $s_1$. 
In effect each point in phase space is a separate oscillation experiment
%% TJG v30
which shows the same oscillator behavior given by Eqn.~\ref{aeqn}. 
In practice this means that one may add together all events of the form
%% TJG B^0 -> B_d
$B_d \to \KS \pi^0 \gamma$ regardless of whether they are produced at the
%% MH v12: K^*(980) --> K^*(892)
$K^*(892)$ resonance, some other resonance ({\it e.g.} $K_2^*(1430)$) 
% $K^*(980)$ resonance, some other resonance ({\it e.g.} $K_2^*(1430)$) 
%% TJG remove continuum
or from a nonresonant source and determine the
%MH26 Sign flip as definition \cals given for 
%MH26 (\bar{\Gamma} - \Gamma)/(\bar{\Gamma} + \Gamma)
single quantity $\cals_{B_d\to\KS \pi^0 \gamma} = +\sin 2\psi^s \sin\phi^s$ 
%MH26 single quantity $\cals_{B_d\to\KS \pi^0 \gamma} = -\sin 2\psi^s \sin\phi^s$ 
as a result. 
Within the SM, 
%MH26 Sign flip as definition \cals given for 
%MH26 (\bar{\Gamma} - \Gamma)/(\bar{\Gamma} + \Gamma)
we obtain $\cals_{B_d\to\KS \pi^0 \gamma} \approx -(2m_s/m_b)\sin 2\phi_1$
%MH26 we obtain $\cals_{B_d\to\KS \pi^0 \gamma} \approx +(2m_s/m_b)\sin 2\phi_1$
%% TJG 2004/11/07 remove comma
as $\psi \approx m_s/m_b$ and
$\phi^s = -2\phi_1 + O(\lambda^2)$, are expected where $\lambda \approx 0.22$
%% TJG 2004/11/07
% is the Cabibbo angle.
is the sine of the Cabibbo angle.
In terms of the individual parameters this gives the lower bounds:

%% MH v12: +\sin...  --> -\sin... Please check it !!
%\begin{eqnarray}
%  \left| \sin 2\psi^s \right| & \geq & \left| S_{\gamma \KS\pi^0} \right|  
% \nonumber\\
%  \left| \sin \phi^s  \right| & \geq & \left| S_{\gamma \KS\pi^0} \right|
%\end{eqnarray}
\begin{eqnarray}
  \left| \sin 2\psi^s \right| & \geq & \left| \cals_{B_d\to\KS \pi^0 \gamma} \right|,
  \nonumber\\
  \left| \sin \phi^s  \right| & \geq & \left| \cals_{B_d\to\KS \pi^0 \gamma} \right|.
  \label{boundeqn}
\end{eqnarray}

%
%
%===============---------->>>>>>>>check this paragraph
%
%

Deviations from this picture of uniform oscillation as a function of phase
space would indicate contributions to the radiative decay other than the
pure dipole transition of the effective Hamiltonian of
%% TJG v30
Eqn.~\ref{effective_H}.

%% TJG B^0 -> B_d
Consider now the case of $B_d \to \pi^+\pi^- \gamma$. 
Again for the $B_d$ decays we can define the Dalitz variables:

\begin{eqnarray}
  s_1 & = & (p_{\pi^+} + p_{\pi^-})^2,     \nonumber\\
  s_2 & = & (p_{\pi^+} + p_{\gamma}  )^2,     \nonumber\\
  s_3 & = & (p_{\gamma}   + p_{\pi^-})^2,     \nonumber\\
  z   & = & \frac{s_3-s_2}{s_3+s_2},
  \label{dalitz2}
\end{eqnarray}

\noindent
and thus write the amplitudes in the form: 
%V27 eq.3 and 4 g_L and g_R need to be switched.  
%\end{eqnarray}
\begin{eqnarray}
  \OL {\cal M}_L(s_1,z) & = & F^{d }_L \OL g_L(s_1,z), \nonumber\\
  \OL {\cal M}_R(s_1,z) & = & F^{d }_R \OL g_R(s_1,z), \nonumber\\
      {\cal M}_R(s_1,z) & = & F^{d*}_L     g_R(s_1,z), \nonumber\\
      {\cal M}_L(s_1,z) & = & F^{d*}_R     g_L(s_1,z).
\end{eqnarray}

\noindent
%V27 footnote on ref. frame added
%% TJG 2004/11/07 correct -> $g_L = - g_R$
As before $g_L = - g_R$
but in this case in applying charge conjugation to
get
from $g_L$ to $\OL g_L$ we need to interchange the coordinates of
the $\pi^+$ and $\pi^-$. 
Thus $s_2 \leftrightarrow s_3$ so 
$z \leftrightarrow -z$, therefore~\cite{ref_frame2}:

%% TJG 2004/11/07 change = to \equiv as per Soni suggestion
%AS25 correcting signs  
\begin{eqnarray}
      g_L(s_1,z) =     - g_R(s_1,z) \equiv - g(s_1, -z),  \nonumber\\ 
  \OL g_L(s_1,z) = - \OL g_R(s_1,z) = - g(s_1,z).
\end{eqnarray}

\noindent 
For particular partial waves of the $\pi^+\pi^-$ system with 
angular momentum $L$, 
%% TJG 2004/11/07 make easier to read
% $g(s_1,z) =  g(s_1,-z)$ for $L$ odd while
% $g(s_1,z) = -g(s_1,-z)$ for $L$ even. 
$g(s_1,z) = (-1)^{L+1} g(s_1,-z)$.
In general $g$ will be a mixture of even and odd $L$, 
%% MH v21: so $f$ has no particular symmetry under $z \leftrightarrow -z$.
so $g$ has no particular symmetry under $z \leftrightarrow -z$.
Note also that $g$ will in general have a nontrivial $CP$-even phase
which
depends on $s_1$ and $z$. 
The time dependent rates 
at a point in phase space defined by $\left( s_1, z \right)$, 
and summed over helicity will be given by:

%% v28 TJG remove \Gamma(t,s_1,z)
\begin{eqnarray}
  \Gamma_{    B_d \to \pi^+\pi^- \gamma}(t,s_1,z) & \propto & 
  e^{-\Gamma t}
  \left[
    X_{\pi^+ \pi^- \gamma}(s_1,z) + Y_{\pi^+ \pi^- \gamma}(s_1,z)\cos(\dmd t) 
    + Z_{\pi^+ \pi^- \gamma}(s_1,z)\sin(\dmd t)
  \right],
  \nonumber\\
  \Gamma_{\OL B_d \to \pi^+\pi^- \gamma}(t,s_1,z) & \propto & 
  e^{-\Gamma t}  
  \left[
    X_{\pi^+ \pi^- \gamma}(s_1,z) - Y_{\pi^+ \pi^- \gamma}(s_1,z)\cos(\dmd t) 
    - Z_{\pi^+ \pi^- \gamma}(s_1,z)\sin(\dmd t) 
  \right].
  \label{genericB}
\end{eqnarray}

Here:
\begin{eqnarray}
  X_{\pi^+ \pi^- \gamma}(s_1,z) & = & (F^d)^2 \left[ |g(s_1,z)|^2+|g(s_1,-z)|^2 
\right],
  \nonumber \\
  Y_{\pi^+ \pi^- \gamma}(s_1,z) & = & (F^d)^2 \left[ |g(s_1,z)|^2-|g(s_1,-z)|^2 
\right],
  \nonumber \\
  Z_{\pi^+ \pi^- \gamma}(s_1,z) & = & - 2 (F^d)^2 
  \left[ 
    \Re \left( g(s_1,-z)g^*(s_1,z) \right) \sin \phi^d + 
    \Im \left( g(s_1,-z)g^*(s_1,z) \right) \cos \phi^d
  \right] \sin 2\psi^d,
  \label{pipi_osc}
\end{eqnarray}

\noindent
and $\phi^d = \phi_M + \phi_L^d + \phi_R^d$ is the weak phase;
%% TJG 2004/11/07 change after discussion with AS
% in the SM $\phi_M = - 2 \phi_1 + O(\lambda^2)$,
% $\phi_L^d = \phi_R^d = \phi_1 + O(\lambda^2)$.  
in the SM $\phi_M \approx - 2 \phi_1$ and $\phi_L^d = \phi_R^d \sim \phi_1$.

%%%%%%%%%%%%%%%%%%%%%%%%%%%%%%%%%%%%%%%%%%%%%%%%%%%%%%%%%%%%%%%%%%%%%%%%%%%%
%%
%% TJG this part moved to straight after \KS\pi^0\gamma section
%%
%%%%%%%%%%%%%%%%%%%%%%%%%%%%%%%%%%%%%%%%%%%%%%%%%%%%%%%%%%%%%%%%%%%%%%%%%%%%
%% Let us now consider the information which can be learned about the terms
%% in the effective Hamiltonian from observing
%% these oscillations.    

%% In the case of $B^0\to \KS\pi^0$, the asymmetry is independent of $s_1$
%% and $z$ so we can combine all the data together. In the end we discover
%% the quantity $a=\sin\phi^s\sin 2\psi^s$. In terms of the individual
%% parameters this gives the lower bounds:

%% \begin{eqnarray}
%% |\sin\phi^s| &\geq& |a|
%% \nonumber\\
%% |\sin2\psi^s| &\geq& |a|
%% \end{eqnarray}

%% The case of $\pi^+\pi^-$ is more complicated. 
%%%%%%%%%%%%%%%%%%%%%%%%%%%%%%%%%%%%%%%%%%%%%%%%%%%%%%%%%%%%%%%%%%%%%%%%%%%%

At each point in phase space, 
%% MH v21: $f(s_1,z)$ and $f(s_1,-z)$ are independent complex numbers. 
%% TJG v21: this is not true if dominated by resonance with known spin
%%          \ldots qualify \ldots
$g(s_1,z)$ and $g(s_1,-z)$ are, in general, independent complex numbers. 
%% TJG v21 Ks\pi^0 \to \KS \pi^0 \gamma
Unlike the case of $\KS \pi^0 \gamma$,
this case does, however, 
allow the possibility of extracting $\phi^d$ and $\psi^d$ separately 
up to discrete ambiguities.
%% TJG v21 Add comment on resonant structure
This can be achieved without any assumption about the 
resonant structure of the $\pi^+\pi^-\gamma$ final state,
as we now demonstrate.
%% TJG qualification
Note that our argument requires regions of phase space where
$\pi^+\pi^-$ partial waves with different angular momentum interfere,
otherwise $g(s_1,z) = \pm g(s_1,-z)$ everywhere and consequently
there is no additional information compared to the $\KS \pi^0 \gamma$ case.

For a given value of $s_1$ and $z$ let us define:

%% TJG v21 s \to s_1
\begin{equation}
  \begin{array}{l@{\hspace{8mm}}l}
    u = \cos 2\psi^d,             &
    a = \frac{|g(s_1,z)|^2-|g(s_1,-z)|^2}{|g(s_1,z)|^2+|g(s_1,-z)|^2}, \\
    v = \sin 2\psi^d \sin \phi^d, &
    b = -2\frac{\Re(g^*(s_1,z)g(s_1,-z))}{|g(s_1,z)|^2+|g(s_1,-z)|^2}, \\
    w = \sin 2\psi^d \cos \phi^d, &
    c = -2\frac{\Im(g^*(s_1,z)g(s_1,-z))}{|g(s_1,z)|^2+|g(s_1,-z)|^2}, \\
  \end{array}
\end{equation}

\noindent
where $a^2 + b^2 + c^2 = u^2 + v^2 + w^2 = 1$. 
%% TJG v21 changed to avoid writing (s_1,z) in the equation
%% The experimental observables in Eqn.~\ref{pipi_osc}
%% are related to these via:
Recalling that the experimental observables,
$X_{\pi^+\pi^-\gamma}$, $Y_{\pi^+\pi^-\gamma}$ and $Z_{\pi^+\pi^-\gamma}$ are functions of the phase space,
we can relate these via:

% \begin{eqnarray}
%   \eta   & = & 
%   \frac{Y_{\pi^+\pi^-\gamma}(s_1,z)}{X_{\pi^+\pi^-\gamma}(s_1,z)} = a
%   \nonumber\\
%   \zeta_+ & = & 
%   \frac{Z_{\pi^+\pi^-\gamma}(s_1,z) + Z_{\pi^+\pi^-\gamma}(s_1,-z)}{2X_{\pi^+\pi^-\gamma}(s_1,z)} = bv
%   \nonumber\\
%   \zeta_- & = & 
%   \frac{Z_{\pi^+\pi^-\gamma}(s_1,z) - Z_{\pi^+\pi^-\gamma}(s_1,-z)}{2X_{\pi^+\pi^-\gamma}(s_1,z)} = cw
% \end{eqnarray}
\begin{eqnarray}
  \eta   & = & 
  \frac{Y_{\pi^+\pi^-\gamma}} {X_{\pi^+\pi^-\gamma}} = a,
  \nonumber\\
  \zeta_+ & = & 
  \frac{Z_{\pi^+\pi^-\gamma} + Z_{\pi^+\pi^-\gamma}}{2X_{\pi^+\pi^-\gamma}} = bv,
  \nonumber\\
  \zeta_- & = & 
  \frac{Z_{\pi^+\pi^-\gamma} - Z_{\pi^+\pi^-\gamma}}{2X_{\pi^+\pi^-\gamma}} = cw.
\end{eqnarray}

%% TJG v21 change limit z \to 0 ==> z = 0 \ldots no reason not to reach this limit
%%In the limit $z \to 0$, $\zeta_-$ and $\eta\to 0$ while $\zeta_+ \to -v$. 
Along the line $z = 0$, we find $\zeta_- = \eta = 0$ while $\zeta_+ = -v$.
Once $v$ is known, we can learn $w$  from cases where $z \neq 0$:

%% MH v22: add \pm explicitly.
\begin{equation}
  w = 
  \frac{\pm\zeta_-}{\sqrt{1-\left(\frac{\zeta_+}{v}\right)^2-\eta^2}} =
  \frac{\pm\zeta_-}{\sqrt{1-\left(\frac{\zeta_+}{\zeta_+(z=0)}\right)^2-\eta^2}}.
\end{equation}

\noindent
From here we can determine $\phi^d$ and $\psi^d$ through

\begin{eqnarray}
  \sin^2 2\psi^d & = & v^2 + w^2, \nonumber\\
%% TJG v21 changing following DA advice
%%  \tan \phi^d    & = & \frac{v}{w}.
  \phi^d & = & \arg \left( (w+iv)/\sin(2\psi^d) \right).
\end{eqnarray}

%{\bf TJG v21: I leave it to someone else to explain the ambiguities.}

%% MH v22: actually we have 16 combinations
%If $(\phi^d,\psi^d)$ is a valid solution, then
%$(\phi^d+\pi^d,\psi^d+\pi^d)$, $(-\phi^d,-\psi^d)$ are also consistent
%with the same observables so there is a 4-fold ambiguity in the
%determination of these angles.
\noindent There are 8 solutions for $\psi^d$. Each of them
has two $\phi^d$ values corresponding to
positive and negative solutions for $w$. Thus there are
16 valid $(\psi^d, \phi^d)$ combinations in total.

In the preceding discussion, we have left the functions
$g_{L,R}$ arbitrary. In practice, the $\pi^+ \pi^-$ amplitude
is very likely to be dominated by low-lying resonances with
well known masses, widths and quantum numbers. This knowledge
could facilitate a more constrained fit to the data. 
In case $B_d\to\rho^0(\to\pi^+\pi^-)\gamma$ is dominant,
we expect $g(s_1,z) =  g(s_1,-z)$ as the $\pi^+\pi^-$
system is in the $L$-odd state.
%% TJG v30
Substituting this into Eqn.~\ref{pipi_osc}, we obtain
$\cals_{B_d\to\rho^0\gamma} = +\sin 2\psi^d \sin 2\phi^d$.

%ANS36 begin

Throughout this discussion we have, again, assumed dominance of the
dipole operators in $H_{\rm eff}$, as in Eqn.~\ref{effective_H}.
To the extent that this assumption holds, $\psi$ and $\phi$ would be
independent of the Dalitz variables, $s_1$ and $z$, as before.  
Conversely, variation of $\psi$ and $\phi$ with $s_1$ and $z$ would 
shed light on the contribution from non-dipole interactions.  
% {\bf MH v12: 
%   Is this true ? tan(-a) = -tan(a) ... 
%   It seems to me that we have 8 solutions for $0 < \psi^d < 2\pi$, 
%   and 2 solutions for $0 < \phi^d < 2\pi$.}

% {\bf TJG v12: 
%   I guess we could also say something about a model dependent
%   fit to the $\pi\pi\gamma$ DP .}

\section{\boldmath $B_s \to K^+ K^- \gamma$ and 
$B_s \to \KS \pi^0(\eta^{(\prime)}) \gamma$}
\label{sectionBsyPP}

Let us consider now the analogous $B_s$ decays. 
In this case we cannot assume that $\Delta \Gamma \approx 0$; 
then the time dependent decay rates to a final state $F\gamma$ are given by:
%AS25, insert s_1,z

%% v28 TJG remove \Gamma(t,s_1,z)
%% TJG 2004/11/07 remove s_1, z again \ldots 
%%                this should follow Eq. 10
\begin{equation}
  \begin{array}{lcr}
    \multicolumn{2}{l}{
      \Gamma_{    B_s \to F\gamma}(t) \propto  
      e^{-\Gamma t}
      \left[
        X_{F\gamma} \cosh(\frac12 \dgs t) + 
        Y_{F\gamma} \cos(\dms t) + 
      \right.
    } & \hspace{50mm} \ \\
    & \multicolumn{2}{r}{
      \left.
        Z_{F\gamma} \sin(\dms t) + 
        W_{F\gamma} \sinh(\frac12 \dgs t)
      \right],
    } \\
    \multicolumn{2}{l}{
      \Gamma_{\OL B_s \to F\gamma}(t) \propto
      e^{-\Gamma t}
      \left[
        X_{F\gamma} \cosh(\frac12 \dgs t) - 
        Y_{F\gamma} \cos(\dms t) - 
      \right.
    } & \\
    & \multicolumn{2}{r}{
      \left.
        Z_{F\gamma} \sin(\dms t) + 
        W_{F\gamma} \sinh(\frac12 \dgs t)
      \right],}
  \end{array}
  \label{genericAs}
\end{equation}

\noindent
%V31, define Gamma in exponent
%% TJG 2004/11/07 remove - \Gamma_L, \Gamma_S are not defined
%% where, in the exponent, $\Gamma = (\Gamma_L + \Gamma_S)/2$ and, 
where

%% MH v22: 
%\begin{eqnarray}
%  X_{F\gamma} & = & |{\cal M}|^2 + |\OL{\cal M}|^2,
%  \nonumber \\
%  Y_{F\gamma} & = & |{\cal M}|^2 - |\OL{\cal M}|^2,
%  \nonumber \\
% Z_{F\gamma} & = & -2 \, \Im \left( e^{i\phi_M} {\cal M}^* \OL{\cal M} \right),
%  \nonumber \\
% W_{F\gamma} & = & -2 \, \Re \left( e^{i\phi_M} {\cal M}^* \OL{\cal M} \right).
%\end{eqnarray}
\begin{eqnarray}
  X_{F\gamma} & = & (|{\cal M}_L|^2 + |{\cal M}_R|^2)
                  + (|\OL{\cal M}_L|^2 + |\OL{\cal M}_R|^2),
  \nonumber \\
  Y_{F\gamma} & = & (|{\cal M}_L|^2 + |{\cal M}_R|^2)
                  - (|\OL{\cal M}_L|^2 + |\OL{\cal M}_R|^2),
  \nonumber \\
  Z_{F\gamma} & = & -2 \, \Im \left( e^{i\phi_M} 
             ({\cal M}_L^* \OL{\cal M}_L + {\cal M}_R^* \OL{\cal M}_R)\right),
  \nonumber \\
  W_{F\gamma} & = & -2 \, \Re \left( e^{i\phi_M} 
             ({\cal M}_L^* \OL{\cal M}_L + {\cal M}_R^* \OL{\cal M}_R)\right).
\end{eqnarray}
%V31, O(lambda^2) not zero 
Here $\phi_M \approx O(\lambda^2)$ is expected for the $B_s$-$\OL B_s$ mixing 
in the SM (in the usual phase convention).

Let us first consider the decay to two mesons which are self-conjugate
such as $B_s\to \KS\pi^0\gamma$.
%AS25 eta has already been said
%needless to say the discussion
%applies equally well when $\pi^0$ is replaced by 
%$\eta^{(\prime)}$.   
%% TJG v30
The expression Eqn.~\ref{genericAs} also describes 
the time dependent decay rates for physical $B_s$ and $\OL B_s$ decays 
at a point in phase space defined by $\left( s_1, z \right)$, 
and summed over photon helicity.
%% TJG again make clear that LD effect neglected
Neglecting long distance effects, 
the factors $X$, $Y$, $Z$ and $W$ are given by

%%MH v21: phi -> phi^d,  psi -> psi^d

%% TJG v21 explicitly add CP of \KS and \pi^0
%AS25 x_i's not needed anymore
%% TJG 2004/11/07 - follow Eq. 22
%%                  \KS\pi^0 should be subscript, (s_1,z) explicit
\begin{eqnarray}
  X_{\KS\pi^0}(s_1,z) & = & 2(F^d)^2 |h(s_1,z)|^2,
  \nonumber\\
  Y_{\KS\pi^0}(s_1,z) & = & 0,
  \nonumber\\
%  Z(\KS\pi^0) & = & 
%  - 2(F^d)^2 \xi_{\KS} \xi_{\pi^0} |h(s_1,z)|^2 \sin 2\psi^d \sin \phi^d,
%  \nonumber\\
%  W(\KS\pi^0) & = & 
%  - 2(F^d)^2 \xi_{\KS} \xi_{\pi^0} |h(s_1,z)|^2 \sin 2\psi^d \cos \phi^d. 
%\end{eqnarray}
  Z_{\KS\pi^0}(s_1,z) & = & - 2(F^d)^2 |h(s_1,z)|^2 \sin 2\psi^d \sin \phi^d,
  \nonumber\\
  W_{\KS\pi^0}(s_1,z) & = & - 2(F^d)^2 |h(s_1,z)|^2 \sin 2\psi^d \cos \phi^d.
\end{eqnarray}

\noindent
In this case, if the value of $W$ can be isolated, both angles $\phi^d$ and
%% TJG v21 add ``up to discrete ambiguities''
$\psi^d$ can be determined, up to discrete ambiguities. 
Of course if $\Delta\Gamma/\Gamma$ is relatively small, this may prove difficult in practice. 
%% TJG B^0 -> B_d
As in the case of $B_d \to \KS \pi^0 \gamma$, 
the asymmetries are independent of $s_1$ and $z$.

In the case of $B_s \to K^+K^-\gamma$, 
the phase space dependent oscillation is again
%% TJG v30
given by the expressions in Eqn.~\ref{genericAs}. 
The $X$, $Y$, $Z$ and $W$ terms are given by:

%% TJG 2004/11/07 - follow Eq. 29
%%                  K^+K^- should be subscript, (s_1,z) explicit
\begin{eqnarray}
  X_{K^+K^-}(s_1,z) & = & (F^s)^2 \left( |d(s_1,z)|^2+|d(s_1,-z)|^2 \right),
  \nonumber\\
  Y_{K^+K^-}(s_1,z) & = & (F^s)^2 \left( |d(s_1,z)|^2-|d(s_1,-z)|^2 \right),
  \nonumber\\
  Z_{K^+K^-}(s_1,z) & = & -2 (F^s)^2 
  \left\{ 
    \Re \left( d(s_1,-z)d^*(s_1,z) \right) \sin \phi^s + 
    \Im \left( d(s_1,-z)d^*(s_1,z) \right) \cos \phi^s
  \right\} \sin 2\psi^s,
  \nonumber \\
  W_{K^+K^-}(s_1,z) & = & -2(F^s)^2 
  \left\{
    \Re \left( d(s_1,-z)d^*(s_1,z) \right) \cos \phi^s - 
    \Im \left( d(s_1,-z)d^*(s_1,z) \right) \sin \phi^s
  \right\} \sin 2\psi^s.
  \label{KK_osc}
\end{eqnarray}

From the $X$, $Y$ and $Z$ terms one can determine $\phi^s$ and $\psi^s$ as
described above in the $B_d \to \pi^+\pi^-\gamma$ case. 
In addition, if the $W$ term can be determined it allows 
for the resolution of the ambiguity between 
$\phi^s$, $\psi^s$ and $\phi^s$, $-\psi^s$.

%A+S2
\section{Analogous Cases, Generalizations and Experimental Considerations}
%A+S2

In the above discussion we have considered some instances of 
%% TJG avoid using subscript q here, which may cause confusion
$B_d \to P_1P_2 \gamma$ and $B_s \to P_1P_2 \gamma$ where 
$P_i$ represent scalar or pseudoscalar states. 
The mode of analysis depends on the nature of $P_1P_2$. 

If there is no relation between $P_1$ and $P_2$ and they are not 
self-conjugate then there will be no oscillations. 
%% TJG B^0 -> B_d
For instance in the case of $B_d \to K^+\pi^-\gamma$ 
we can tell from the final state whether the initial state is $B$ or $\OL B$ 
so no quantum interference is possible.
Final states of this sort do however give the simplest way to determine if
there is direct $CP$ violation at the quark level.

If $P_1$ and $P_2$ are both eigenstates of charge conjugation,
%% TJG B^0 -> B_d
then the mode of analysis is the same as $B_d \to \KS\pi^0 \gamma$. 
%% TJG should be \sin 2\psi^q \sin \phi^q (?)
In this case we can learn the product $\sin 2\psi^q \sin \phi^q$,
%% TJG make meaning of superscript explicit
where the superscript $q$ represents the quark produced in $b \to q \gamma$. 
For $B_s$ oscillations, if the $W$ term can be extracted 
then we can learn $\phi^q$ and $\psi^q$ separately. 
In all cases, the data can be integrated over $s_1$ and $z$. 
%%MH v21: Some other final states of this type are $\KS \eta^{(\prime)} \gamma$, $\pi^0 
Some other final states of this type, 
%v31...must say B_d as they are not for B_s
relevant to $B_d$ decays,
are $\KS \eta^{(\prime)} \gamma$,
$\KS f_0\gamma$, $\pi^0 \eta^{(\prime)} \gamma$, $\pi^0 f_0 \gamma$, 
%% TJG 2004/11/07 add emphasis on \KS \eta^\prime \gamma 
For most of these modes there is currently no experimental information;
they are therefore in urgent need of investigation.
%v31
Cases that may be of special interest are $B^0 \to K_S \eta^{\prime} \gamma$
and $B^0 \to K_S \eta \gamma$. 
% Since $\frac{Br(B \to K_S \eta^{\prime})}{Br(B\to K_S \pi^0)} \approx 7$, 
% it is plausible that $B \to K_S \eta^{\prime} \gamma$ is also more copious than 
% $B \to\gamma [K_S \pi^0]_{\rm nonresonant}$.
%are $\KS \eta^{(\prime)} \gamma$,
%$\KS f_0\gamma$, $\pi^0 \eta^{(\prime)} \gamma$, $\pi^0 f_0 \gamma$,
%% TJG 2004/11/07 add emphasis on 
%v31
%However in some cases, such as $\KS \eta^\prime \gamma$ 
Comparisons to the pattern of branching fractions in two body hadronic
$B$ decay suggest that the nonresonant contribution to 
$B \to K \eta^\prime \gamma$ ought to be larger than that for either 
$B \to K \eta \gamma$ or $B \to K \pi^0 \gamma$.
Note also that $B \to K \eta \gamma$ has recently been observed
with branching ratio $\approx 7 \times 10^{-6}$~\cite{bc0412}.
% Naively, we expect the branching ratio for  $B \to K \eta^\prime \gamma$ 
% ought to be larger than that for $B \to K \eta \gamma$.
Therefore it is possible that an appreciable data sample for 
$\KS \eta^\prime \gamma$
%it is possible that fairly large samples 
may already be available with the current $B$ factory statistics.  

%% MH include f_0 ?
%
%%MH v21: I think that this is not another case, but is a special case of
% both P1 and P2 being C eigenstates.
%% TJG v21 I agree, and changed it
%%Another case for  which the same comments apply to is when $P_1 = P_2$, 
Note that there is a set of special cases of these modes, where $P_1 = P_2$, 
for instance, $\KS\KS \gamma$ and $\pi^0\pi^0 \gamma$ 
(the latter unlikely to be of practical use unless
a very high luminosity allows us to 
use $\pi^0\to e^+e^-\gamma$ decays that provide vertex information). 
In these cases, Bose-Einstein statistics further constrain the 
$P_1 P_2$ system~\cite{GH}.

%% MH v21: Discussions on the more general n-body case
%% TJG v21 English corrections
%% V30 MH As explained in Sec.~\ref{sectionByPP}, 
As explained in Sec.~\ref{sectionByPP}, 
the relative angular momentum between $P_1$ and $P_2$ does not enter;
%% v28 TJG updated
%% TJG 2004/11/07 $C$ parity -> charge conjugation quantum number
only the intrinsic 
charge conjugation quantum numbers
of $P_1$ and $P_2$ affect $CP$ asymmetries.
%only the intrinsic eigenvalues of $P_1$ and $P_2$ affect the asymmetries.
This is also valid in more general cases with more than two particles,
{\it e.g.} $B\to P_1 P_2 P_3\gamma$, where $P_1$, $P_2$ and $P_3$ are
eigenstates of charge conjugation. 
%AS25 add some specific examples
Specific examples of this type are $B \to K_S \pi^0 \pi^0 \gamma$,
%% TJG 2004/11/07 - change \ldots to and so on
$K_S K_S K_S \gamma$, $K_S K_S \eta^{(\prime)} \gamma$, and so on.
%A+S2
%% TJG 2004/11/07 - cut
% One such case that may be of special interest is $B^0 \to K_S
% \eta^{\prime}(\eta) \gamma$. Since $\frac{Br(B \to K_S \eta^{\prime})}{Br(B
% \to K_S \pi^0)} \approx 7$, it is plausible that $B \to K_S \eta^{\prime}
% \gamma$ is also more copious than $B \to K_S \pi^0 \gamma$. 

%
%
%%%>>DA3
%
%
Indeed, as long as all the final state particles are eigenstates of
charge conjugation, regardless of their spin or other quantum numbers,
%% TJG v30
Eqn.~\ref{aeqn} applies since charge conjugation does not map one point of
phase space to another.  A case of particular interest is $B^0\to K_S \phi \gamma$. 
This final state has the practical advantage that the $\phi$ can
be observed as $K^+K^-$,
%% TJG 2004/11/07 - improve readability
allowing a simple determination of the $B$ decay vertex,
rather than the extrapolation which is necessary when using $K_S\to\pi^+\pi^-$. 
%% TJG 2004/11/07 - improve readability
Note that there is an indication that this mode has a branching fraction
of $\sim 5 \times 10^{-6}$~\cite{belle_phikg}.
% also this mode has been seen~\cite{belle_phikg},
% %% TJG 2004/11/07 - less precise BF
% with a branching ratio $\sim 5 \times 10^{-6}$.  
%v31 ...must emphasize need of a firm resonance cut in this case
Other analogous cases of interest are $K_S \rho \gamma$ and $K_S \omega \gamma$.
% In all such cases ({\it i.e.} $K_S V \gamma$) it would be essential
% to include only the resonant ($V = \phi$, $\rho$, $\omega$) contributions. 
%
%
%%%>>DA3
%A+S2
%
%

If $P_1$ and $P_2$ are not self-conjugate but are anti-particles of each
%% TJG v21 $\pi^+\pi^-$ \to $\pi^+\pi^- \gamma$
other then the mode of analysis is the same as $\pi^+\pi^- \gamma$ above. 
In this case by considering the time dependent Dalitz plot we can separately
determine $\phi^q$ and $\psi^q$. 
Final states of this sort are $\pi^+\pi^- \gamma$ and $K^+K^- \gamma$. 
Table~\ref{tableA} shows the various final states which may result from 
$B_d$ and $B_s$ decay and which quark-level  process they are sensitive to.

\begin{table}
  \begin{tabular}
    {|@{\hspace{5mm}}c@{\hspace{5mm}}|@{\hspace{5mm}}c@{\hspace{5mm}}|
      @{\hspace{5mm}}c@{\hspace{5mm}}|@{\hspace{5mm}}c@{\hspace{5mm}}|
      @{\hspace{5mm}}c@{\hspace{5mm}}|@{\hspace{5mm}}c@{\hspace{5mm}}|}
%\cline{2-6}
%\multicolumn{1}{c}{ }     & \multicolumn{5}{|c|}{Final State}\\
%     & $\KS\pi^0$ (or $\eta^{(\prime)})$ & $\KS\KS$ & $\pi^+\pi^-$ & $K^+K^-$ & 
%\KS\KL$ \\
\hline

& $\KS\pi^0\gamma$ & $\KS\KS\gamma$ & $\pi^+\pi^-\gamma$ & $K^+K^-\gamma$ & $\KS\KL\gamma$ \\
    \hline
    \hline
    $B_d / \OL B_d$ & $b\to s\gamma$ & $b\to d\gamma$ & $b\to d\gamma$ & $b\to d\gamma$ & $b\to d\gamma$ \\
    \hline
    $B_s / \OL B_s$ & $b\to d\gamma$ & $b\to s\gamma$ & $b\to s\gamma$ & $b\to s\gamma$ & $b\to s\gamma$ \\
    \hline
  \end{tabular}
  \caption{
    Final states which can be used to probe $b \to s \gamma$ and $b \to d \gamma$ 
    transitions in $B_d$ and $B_s$ decays.
    This list is not exhaustive; in particular other neutral (pseudo-)scalar
    particles ($\eta$, $\eta^\prime$, $f_0$) may be used in the place of $\pi^0$.
  }\label{tableA}
\end{table}

Let us now consider a few of these cases which are likely to be of
greatest experimental interest. 
%% TJG B^0 -> B_d
In the case of $B_d$ decay preliminary
studies have been done of the AGS mode $\KS\pi^0 \gamma$ on the 
$K^*$ resonance~\cite{expt_ags}.
Part of the importance of the discussion 
%above 
in this paper
is that this data may be
combined with nonresonant $\KS\pi^0\gamma$ decays
%AS25 add line below
and also with signal from other resonances. 
Since we are interested in the underlying two-body process $b \to q \gamma$,
we assume that experimental cuts will be imposed to 
discriminate against bremsstrahlung
%AS25 add line below
and other possible background.
At a $B$-factory experiment, these cuts will typically include a requirement 
on the center-of-mass energy of the photon;
since the $B$ meson is almost at rest in the $\Upsilon(4S)$ rest frame
this is equivalent to requiring a hard photon and will remove 
%% v28 TJG most of -> most
most bremsstrahlung events.
Further cuts may include other requirements on the 
$P_1 P_2 \gamma$ phase space.
In addition to reducing the dominant experimental background from
continuum $e^+e^- \to q\OL{q}$, $q = u,d,s,c$ events,
these requirements can remove the background 
caused by two body hadronic $B$ decays followed by radiative 
hadronic decays.
For example, the decay chain $B_d \to \omega \KS$, $\omega \to \pi^0 \gamma$,
contributes to the final state $\KS \pi^0 \gamma$.
This background decay has a product branching fraction of 
$2.5 \times 10^{-7}$~\cite{pdg},
which is small, but not entirely negligible compared to that for the signal.
However, this background can be removed with a requirement on the
invariant mass of the $\pi^0\gamma$ combination.
Similar backgrounds should be considered for each final state.

Replacing the $\pi^0$ with an $\eta^{(\prime)}$ is more challenging 
experimentally,
%V30 MH although the decay $B_d \to K \eta \gamma$ has recently been 
although the decay $B \to K \eta \gamma$ has recently been 
observed~\cite{bc0412}. 
These decays should measure the same quantity if the decay is controlled by
%% TJG v30
Eqn.~\ref{effective_H}. 
It would however be of some importance to use these modes as a 
check that new physics proceeds through this dipole operator.

In the $B_d$ system $\KS\KS \gamma$ monitors $\sin \phi^d \sin 2\psi^d$ for the
%% TJG this should be b->d, I guess
%%$b\to s\gamma$ transition. 
$b\to d\gamma$ transition. 
Of course this case may also be subject to a
significant amount of direct $CP$ violation at the quark level which 
%% TJG can get DCPV from time (in)dependent analysis
can also be measured in the usual way in a time (in)dependent analysis,
and could be further checked via charged $B$ decays such as 
$B^+ \to K^+\KS \gamma$ or $B^+ \to \pi^+\pi^0\gamma$.

%% TJG v21 add B_d to make explicit for these decays
The $P^+P^- \gamma$ states $B_d \to \pi^+\pi^-\gamma$ and $B_d \to K^+K^-\gamma$ could
in principle be used to separately obtain $\phi^d$ and $\psi^d$. 
The $B_d \to K^+K^- \gamma$ case is, however,
%% TJG change Zweig -> OZI because MH likes it more
%% Zweig
OZI suppressed and so is unlikely to have a significant branching ratio. 
%% TJG add $B_s \to \pi^+\pi^- \gamma$
The same is true for $B_s \to \pi^+\pi^- \gamma$.
Experimentally, good $K/\pi$ separation is needed to isolate 
%% TJG B^0 -> B_d
$B_d \to \pi^+\pi^-\gamma$ from the decay $B_d \to K^+\pi^-\gamma$ 
which may have an order of magnitude greater branching ratio.

Observing $B_s$ oscillations at hadronic $B$ machines will of course be
challenging. 
The simplest of the modes in Table~\ref{tableA} to measure is probably 
$B_s \to K^+K^-\gamma$ which again is sensitive to the $b \to s\gamma$ 
transition. 

%\section{Standard Model Expectations}
\section{Annihilation Contribution} 

The key feature of oscillations in $B_q \to F\gamma$ is that they only may
take place if $\psi \neq 0$. 
In the SM,
this reaction takes place through the penguin in Fig.~\ref{feydia_a}.  
The photon due to this penguin process is exactly left handed 
in the limit of massless $q$; since in that case $q$ would be left handed, 
so conservation of helicity would imply that the the photon 
would likewise necessarily be left handed. 
More generally, in the SM this process gives rise to $\psi \approx m_q/m_b$.

Another source of right handed photons which could therefore potentially
produce a signal is the annihilation diagram shown in
Fig.~\ref{feydia_d}. 
This process would only 
%AS25 modeify a bit
contribute as a background to
be present in 
$b \to d\gamma$ processes in 
$B_d$ decays and to $b \to s\gamma$ processes in $B_s$ decays. 
These processes are subject to large nonperturbative effects when
the photon is radiated  colinearly with the 
%virtual 
initial light quark. 
This enhancement, however, is only in the left handed photon channel. 
To see this consider the amplitude of the relevant 
annihilation graph (d) where we
have applied a Fierz transformation to the $W$ propagator:

\begin{eqnarray}
  {\cal M}_{\rm ann} & \propto &
  \frac{1}{p_d^2-2q\cdot p_d}
  \left ( \OL d \slE (\slq-\slp_d) \gamma^\mu L b\right )
  \left ( \OL d  \gamma_\mu L u \right ),
\end{eqnarray}

\noindent
where $p_d$ is the momentum of the initial $d$ quark,
$L=(1-\gamma^5)/2$ and $R=(1+\gamma^5)/2$. We can rewrite this
in the following form:

%{\bf TJG v21: surely something must be wrong here?}
%AS25, yes a plus sign in between two terms was missing! 

\begin{eqnarray}
  {\cal M}_{\rm ann} & \propto &
  -i\frac{f_B}{p_d^2-2q\cdot p_d}
  \left ( \OL d  \left[ \sigma^{\mu\nu}E_\mu q_\nu R\right ] 
    \gamma^\mu b\right )
  \left ( \OL d  \gamma_\mu L u \right )
%   \nonumber\\
%   &-&
+
  \frac{f_B}{p_d^2-2q\cdot p_d}
  \left ( \OL d \slE \slp_d \gamma^\mu L b\right )
  \left ( \OL d  \gamma_\mu L u \right ).
\label{annh}
\end{eqnarray}

%AS25 add line below
%% TJG 2004/11/07 B -> $B$
\noindent where $f_B$ is the $B$ decay constant and $E_\mu$ is the polarization
4-vector of the photon.  
The first term is by itself gauge invariant. 
The factor in square brackets is the dipole operator 
for the emission of left handed photons. 
This term is enhanced by the propagator in front because 
this quantity in the rest frame of the $B$ meson is approximately 
$1/(E_\gamma E_d)$ where $E_d$ is expected to be small
%AS25 add lambda. 
$\approx \lambda_{QCD}$. 
As discussed in~\cite{bloc} this term corresponds to emission of a 
colinear photon by a light (initial) 
quark and because of its singular nature 
%% TJG likely -> probably
in perturbation theory it is expected to receive non-perturbative corrections. 
Since it has the same photon helicity as the penguin produced
photons but a different CKM phase, 
it will alter the magnitude and phase of $F_L^q$ 
but will not contribute to $F_R^q$ and therefore
%% v28 TJG changed 
%%will not contribute to $\psi^q$.
will not much affect $\psi^q$.

%{\bf TJG v21: should this be ``will not significantly affect $\psi^q$''?
%  surely $\tan \psi^q = F_L^q/F_R^q$?}
%AS25 Actually, tan psi^q = F_R/F_L so in the SM it is very small.
%AS25 what the above discussion says is that the denominator of
%F_R/F_L will change a bit but will not alter conclusions
%in any significant way 

%ref(begin)  

Actually, the singular nature of the first term in 
Eqn.~\ref{annh} is a consequence of the very simple non-relativistic
model used for the bound state. Indeed in more sophisticated
discussions, that quantity is not singular anymore but has a well
defined light-cone expansion in $1/E_\gamma$~\cite{new_ref}; 
the important point relevant to this paper is that the left-handed
nature of the simplistic picture above survives the improved
theoretical treatment. 

%ref(end)

The second term is not gauge invariant but must form a gauge invariant set
%% TJG graphs -> diagrams
when added to the other bremsstrahlung diagrams. 
%% TJG 2004/11/07 proportional to
This term is proportional to the 4-momentum of the light quark 
and so the energy of the light quark in the numerator will tend to cancel the
denominator. 
Therefore, though these graphs will produce photons of both helicity 
they are not enhanced 
by a light quark propagator and so are expected to make
only a negligible contribution.

Thus we conclude that the annihilation
contribution does not cause any serious difficulty 
to the analysis above.

%
%
%ANS36 begin
%DA35(4)
%
\section{Non-Dipole Operators}
\label{non-dipole}

A potential complication for the use of this method as a test for new
physics is the SM contribution to radiative decays through non-dipole
operators~\cite{grinstein}, 
for instance, the penguin $b\to s \gamma g$. 
Such processes generally do
not fix the helicity of the photon and so would result in a SM
contribution to ${\cal S}$.
%ASv38...small wording chnage
Also the precise contribution of
such processes to particular final states is difficult to
calculate reliably. Fortunately,
the experiments being suggested here
can give a handle on the extent of these non-dipole
contributions 
%ASv38
and the presence of the latter need not
represent a serious limitation to the application
of our method to search of NP. 
%ASv38

The different operator structure in $H_{\rm eff}$ would mean, that 
in contrast to
the pure dipole case, ${\cal S}$ would depend on the kinematics and
composition of the final state. 
For example, for the case of modes such as $K_S \pi^0 (\eta, \eta^{\prime}) \gamma$ 
(see the discussion after Eqns.~\ref{aeqn}~and~\ref{boundeqn}) 
the presence of non-dipole contributions would, in general,
make the asymmetry ${\cal S}$
depend on the Dalitz variables $s_1$ and $z$.
%ASv38-
Thus, the contamination from non-dipole terms in $H_{\rm eff}$ may be
estimated by fitting the experimental data on ${\cal S}$ to
%ASv38 i think we should be very explicit...from a polynomial everyone
%can readily see how the cosntant can be retrieved.  
a suitable parametrization of the dependence on $s_1$ and/or $z$.
%ASv38 begin 
%  The value of ${\cal S}$ that is independent of
%  $s_1$ may then be identified with the dipole interaction.
%ASv38 ..end
A difference in the values of
${\cal S}$ for two resonances of identical $J^{PC}$ would also
indicate non-dipole contributions.  
Similarly for the $\pi^+ \pi^- \gamma$ mode,
$\psi$ and $\phi$ would depend on $s_1$ and z. 
For a better understanding of these non-leading effects, it is
also useful to measure ${\cal S}$ for a number of different final
states and kinematic ranges and see if there is variation. For instance,
in the case of $b\to s \gamma$ some decays which should be studied are
those suggested above of the form $B^0\to \gamma K_S + {\rm neutrals}$ where
${\rm neutrals}$ could include $\pi^0$, $\eta^{(\prime)}$, $\rho^0$, $\omega$,
$\phi$, $K_S$, $\dots$ or any combination of such mesons.

%ANS36 end

\section{Summary}

We have extended the work of~\cite{ags1} so that data from
other resonances as well as nonresonant contributions to final
%% TJG v21 changed \gamma P1P2 to P1P2 \gamma for consistency
states such as $\KS \pi^0 \gamma$, $\KS f_0 \gamma$, $K^+ K^- \gamma$,
%% TJG v21 changed K^0 \OL K^0 \to \KS\KS
%% TJG 2004/11/07 $\KS \eta (\eta') \gamma$ -> $\KS \eta^{(')} \gamma$
$\pi^+ \pi^- \gamma$, $\KS \eta^{(')} \gamma$ and $\KS \KS \gamma$
can be
included for the $b \to s$ or $b \to d$ transitions relevant to
%AS25 KSKS needs a footnote
$B_d$ or $B_s$ decays.
%% TJG 2004/11/07  this citation not necessary ~\cite{ksks}.  
This should significantly improve the effectiveness of
testing the Standard Model with mixing-induced $CP$ violation
in radiative $B$ decays. 
%% TJG 2004/11/07 not only this -> this not only
Indeed, this not only helps in
reducing the statistical errors, since no separation 
between resonant and nonresonant events is
needed, it should also help in reducing the systematic errors.
%% TJG 2004/11/07 - don't need to give list every time
%% For $\KS \pi^0 (\eta, \eta') \gamma$ type of 
For the $\KS \pi^0 \gamma$ type of 
final state an improved determination
of the product $\sin 2 \psi^q \sin \phi^q$, where $\phi^q$ is
the weak phase and $\psi^q$ monitors
the photon helicity, is possible.
%% TJG v21 removed K^0 \OL K^0 \gamma
For final states such as $\pi^+ \pi^- \gamma$ and $K^+ K^- \gamma$,
separate determination of each of these two quantities is possible.

%ASv38
A key feature of these modes is that the dipole interaction
in $H_{\rm eff}$ gives rise to a mixing-induced CP asymmetry
which is independent of the Dalitz variables. Recall that
in many models of New Physics it is the (dimension five) dipole
interaction that is likely to be dominant. Thus the study
of the CP asymmetries of these modes should be useful
in searching for New Physics even in the presence of 
non-dipole SM contributions.
%ASv38

In passing, we briefly recall the hierarchy of $CP$ asymmetries in 
radiative $B$ decays expected in the Standard Model. 
%% TJG 2004/11/07 - improve readability
%ANS 36 begin
Assuming the dipole Hamiltonian dominates,
%ANS 36 end
for $b \to s$ the mixing-induced $CP$ asymmetry is expected to be 
$O(3\%)$ and the direct
$CP$ asymmetry~\cite{dirCP} should be around $0.6\%$. For $b \to d$, the direct
%%MH v21: $CP$ asymmetry is expected to be around $16\%$ whereas the mixing-induced
$CP$ asymmetry is expected to be around $15\%$ whereas the mixing-induced
$CP$ asymmetry is $\approx 0.1\%$, making it into a very
interesting (essentially) {\it null test} of the SM.

\section*{Acknowledgments}

%v31
%ANS36 begin  
We want to thank Yuval Grossman and Alex Kagan for discussions.  
%ANS36 end
This work grew from discussions at the Super $B$ Factory Workshop
held at University of Hawaii (Jan. '04). A. S. thanks the members
of the Belle Collaboration for their 
hospitality during his visit to KEK for completion of this work. 
T.~G. is supported by the Japan Society for the Promotion of Science.
The work of D.~A. and A.~S. are supported in part by US DOE grant Nos.
DE-FG02-94ER40817 (ISU) and DE-AC02-98CH10886 (BNL).

\section*{Appendix: The Effect of a Perturbative Phase}

As discussed in~\cite{dirCP} the $b\to d\gamma$ transition
is expected to receive contributions with different weak and 
strong phases which could
lead to appreciable direct $CP$ violation  at the quark level. 
%% TJG B^0 -> B_d
If we take this effect into account in the case of $B_d \to \pi^+\pi^-\gamma$,
the modified values of $X$, $Y$ and $Z$ are:

%% MH v21: f --> g
%% TJG v21 made this prettier and fixed hspace warning
\begin{equation}
  \begin{array}{lc@{\hspace{30mm}}r}
    \multicolumn{2}{l}{
      X(\pi^+\pi^-\gamma) = F^2 
      \left\{
        \left[ |g(s_1,z)|^2 + |g(s_1,-z)|^2 \right] (1+\delta^2) + 
        2 \delta \cos^2\psi \left[ \cos(\mu-\nu) g(s_1,z) + \cos(\mu+\nu) g(s_1,-z) 
        \right]
      \right\},
    } &
    \\
    \multicolumn{2}{l}{
      Y(\pi^+\pi^-\gamma) = F^2
      \left\{ 
        \left[ |g(s_1,z)|^2 - |g(s_1,-z)|^2 \right] (1+2\delta\cos^2\psi \cos\mu \cos\nu) -
      \right.
    } & 
    \\
    & \multicolumn{2}{r}{
      \left.    
        \left[ |g(s_1,z)|^2 + |g(s_1,-z)|^2 \right] 2 \delta\cos^2\psi \sin\mu\sin\nu
      \right\},
    }
    \\
    \multicolumn{2}{l}{
      Z(\pi^+\pi^-\gamma) = -2(F^d)^2 
      \left\{ 
        \Re \left( g(s_1,-z)g^*(s_1,z) \right) 
        \left[ \sin\phi^d + \delta \cos\mu \sin(\phi^d+\nu) \right] +
      \right.
    } &
    \\
    & \multicolumn{2}{r}{
      \left.
        \Im \left( g(s_1,-z)g^*(s_1,z) \right) 
        \left[ \cos\phi^d \delta \cos\mu \cos(\phi^d+\nu) \right]
      \right\} \sin 2\psi^d.
    }
  \end{array}
\end{equation}

% \begin{eqnarray}
%   X(\pi^+\pi^-\gamma) & = & F^2 
%   \left\{
%     \left[ |g(s_1,z)|^2 + |g(s_1,-z)|^2 \right] (1+\delta^2) + 
%     2 \delta \cos^2\psi \left[ \cos(\mu-\nu) g(s_1,z) + \cos(\mu+\nu) g(s_1,-z) 
% \right]
%   \right\}
%   \nonumber\\
%   Y(\pi^+\pi^-\gamma) & = & F^2
%   \left\{ 
%     \left[ |g(s_1,z)|^2 - |g(s_1,-z)|^2 \right] (1+2\delta\cos^2\psi \cos\mu 
% \cos\nu) -
%     \left[ |g(s_1,z)|^2 + |g(s_1,-z)|^2 \right] 2 \delta\cos^2\psi 
% \sin\mu\sin\nu
%   \right\}
%   \nonumber\\
%   Z(\pi^+\pi^-\gamma) & = & -2(F^d)^2 
%   \left\{ 
%     \Re \left( g(s_1,-z)g^*(s_1,z) \right) 
%     \left[ \sin\phi^d + \delta \cos\mu \sin(\phi^d+\nu) \right]
%   \right.
%   \nonumber\\
%   & &
%   \left.
%     + \Im \left( g(s_1,-z)g^*(s_1,z) \right) 
%     \left[ \cos\phi^d \delta \cos\mu \cos(\phi^d+\nu) \right]
%   \right\} \sin 2\psi^d.
% \end{eqnarray}


\begin{thebibliography}{99}
\bibitem{abc}
We use the $\left( \phi_1, \phi_2, \phi_3 \right)$ notation for the angles of 
the Unitarity Triangle; recall: $(\phi_1, \phi_2, \phi_3) \equiv 
(\beta, \alpha, \gamma)$.
\bibitem{asp_def}
Note that the definition of the $CP$ asymmetry used here follows the
convention of~\cite{pdg_review} and is opposite to that used
in~\cite{ags1}; see also~\cite{abc}.
\bibitem{c_not_cp}
Note another change from~\cite{ags1};
in that paper it was stated 
%v31 soften a bit as its not needed
%incorrectly
that $CP$ of the resonance $V$ ($K^*$ or $\rho$) 
determines the sign of oscillations;
%v31
in practice for $V$ case it makes no essential difference.
\bibitem{also_eta} Throughout our theoretical
discussion, $\pi^0$ can always be replaced by $\eta$ or $\eta^{\prime}$
or any other non-strange scalar.
\bibitem{ref_frame1}
%% v28 TJG English corrections
We have used a reference frame where the $B$ is at rest with the decay
occurring in the $xz$ plane, the photon momentum along the $-z$ axis and the
$x$ component of the $K_S$ momentum is positive. If we use a more general
coordinate system where the photon momentum is in the $-z$ direction,
then we need to introduce the azimuthal angle, $\phi$, of the $K_S$ momentum. 
In this case, ${\cal M}_{^R_L}(s_1,z)=\pm F_{^L_R}f_(s_1,z)e^{\pm i\phi}$.
\bibitem{ref_frame2}
%% v28 TJG English corrections
As in the previous case, we have used a reference frame where the $B$ is
at rest with the decay occurring in the $xz$ plane, the photon momentum along
the $-z$ axis and the $x$ component of the $\pi^+$ momentum is positive.
In the more general case where the photon momentum is in the $-z$
direction, then we need to introduce the azimuthal angle, $\phi$, of the $\pi^+$. 
In this case, $\OL{\cal M}_{^R_L}(s_1,z)= \pm F_{^R_L}g(s_1,z)e^{\pm i\phi}$. 
Note however that the application of charge conjugation on the
%%V30 MH phase space ($z$, $\phi$) in Eqn.~{\ref{MShortDistance}} gives ${\cal
phase space ($z$, $\phi$) in Eqn.~\ref{MShortDistance} gives ${\cal
M}_{^R_L}(s_1,z)$ = $\pm F_{^L_R}^*g(s_1,-z)e^{\pm i(\phi+\pi)}
C_{\rm internal}$ = $\pm F_{^L_R}^*g(s_1,-z)e^{\pm i\phi}$ since
$C_{\rm internal}=C_\gamma=-1$ while the change $\phi\to\phi+\pi$ due to
interchange of $\pi^+\leftrightarrow \pi^-$ also introduces a factor of
(-1).
\end{thebibliography}
\end{document}